\PassOptionsToPackage{usenames,dvipsnames}{xcolor}
\documentclass[11pt,letterpaper]{article}

\usepackage[T1]{fontenc}
\usepackage[latin9]{inputenc}
\usepackage[authoryear]{natbib}
\usepackage[caption=false]{subfig}
\usepackage{lscape}
\usepackage{multirow}
\usepackage{ae,aecompl,geometry,float,framed,rotfloat,amsthm,soul,amsmath,
amssymb,graphicx,todonotes,xcolor,setspace,enumitem,sectsty, colortbl,
booktabs,rotating,microtype,soul,units,bbm,mathtools,tipa,fancybox,makecell,palatino}

\usepackage{hyperref}
  \hypersetup{colorlinks=true,citecolor=Bittersweet,linkcolor=Bittersweet,urlcolor=Bittersweet}
\makeatletter

\theoremstyle{plain}

  \theoremstyle{plain}
  
  \theoremstyle{definition}
  
  \theoremstyle{plain}
  
  \theoremstyle{plain}
  
  \theoremstyle{plain}
  \newtheorem*{cor*}{\protect\corollaryname}
\newcounter{asscount}
\newtheorem{hypth}{Hypothesis}

\newtheoremstyle{assumption}
  {0.2cm}{0cm}
  {\rmfamily}
  {0cm}
  {\bfseries}{ }
  {0cm}
  {\thmname{#1}\thmnumber{ #2}:\thmnote{ #3}}
\theoremstyle{assumption}

\def\!{\mskip-\thickmuskip}
\def\?{\!\!\!}

\definecolor{LightCyan}{rgb}{0.615, 0.815, 0.984}

\@ifundefined{showcaptionsetup}{}{%
 \PassOptionsToPackage{caption=false}{subfig}}
\usepackage{subfig}
\makeatother
  \providecommand{\corollaryname}{Corollary}
  \providecommand{\definitionname}{Definition}
  \providecommand{\lemmaname}{Lemma}
  \providecommand{\propositionname}{Proposition}
\providecommand{\theoremname}{Theorem}

\usepackage[framemethod=tikz]{mdframed}
\usetikzlibrary{shadows}
\newmdenv[shadow=true,shadowcolor=black,rightmargin=8pt]{shadedbox}

\usepackage{titling}

\begin{document}

\newcommand{\titlepaper}{Do speed bumps curb low-latency trading? \\ Evidence from a laboratory market}

\onehalfspacing  
\title{\textbf \titlepaper}

\newcommand{\mazabstract}{\noindent Exchanges implement intentional trade delays to limit the harmful impact of low-latency trading. Do such ``speed bumps'' curb investment in fast trading technology? Data is scarce since trading technologies are proprietary. We build an experimental trading platform where participants face speed bumps and can invest in fast trading technology. We find that asymmetric speed bumps, on average, reduce investment in speed by only 20\%. Increasing the magnitude of the speed bump by one standard deviation further reduces low-latency investment by 8.33\%. Finally, introducing a symmetric speed bump leads to the same investment level as no speed bump at all. 
\bigskip{}
}

\author{Mariana Khapko and  Marius Zoican\thanks{Mariana Khapko is affiliated with University of Toronto Scarborough and  the Rotman School of Management. Marius Zoican is affiliated with University of Toronto Mississauga and the Rotman School of Management. Mariana Khapko can be contacted at \href{mailto:Mariana.Khapko@Rotman.Utoronto.Ca}{mariana.khapko@rotman.utoronto.ca}. Marius Zoican can be contacted at \href{marius.zoican@rotman.utoronto.ca}{marius.zoican@rotman.utoronto.ca}. The authors are grateful to Philipp Chapkovski for valuable research assistance, and to Eric Aldrich, Michael Brolley, David Cimon, Marie-Pierre Dargnies, Michele Dathan, Marlene Haas, Charles Martineau, Sophie Moinas, and Andriy Shkilko for helpful discussions on this research, as well as to Catherine Boullianne and Bill Chau for logistical support. Mariana Khapko and Marius Zoican gratefully acknowledge the Canadian Social Sciences and Humanities Research Council (SSHRC) for an Insight Development research grant.}}

\maketitle

\vspace{-3mm}


\begin{abstract}
\mazabstract 

\noindent \textbf{Keywords}: high-frequency trading, experimental finance, speed bumps, trading technology

\noindent \textbf{JEL Codes}: C90, G11, G14, G40

\thispagestyle{empty}

\newpage{}
\thispagestyle{empty}
\end{abstract}

\vfill{}

\vfill{}

\pagebreak{}

 \thispagestyle{empty}

\vspace*{20mm}
\begin{center}
\huge \titlepaper
\par\end{center}{\Large \par}

\vspace{12mm}

\bigskip{}
\begin{abstract}
\mazabstract

\bigskip{}

\textbf{Keywords}: high-frequency trading, experimental finance, speed bumps, trading technology

\textbf{JEL Codes}: C90, G11, G14, G40
\bigskip{}

\newpage{}
\setcounter{page}{1}
\end{abstract}













\newpage
\setcounter{page}{1}

\section{Introduction \label{sec:Introduction}}

Does a slower exchange reduce individual traders' appetite for speed? After years of leveraging cutting-edge technology to offer ever-faster trading to investors \citep{PagnottaPhilippon2018}, exchanges have started to slow down -- rather than speed up -- the market. The underlying rationale for this slow-down is that high-frequency traders, with the ability to submit orders at almost the speed of light, have been held responsible for earning profits at the expense of slower traders and crowding them out of fast exchanges.

Most efforts to slow down order traffic materialize as \emph{speed bumps}: that is, intentional delays between the receipt and execution of (a subset of) trading orders submitted to the exchange. The length and design of the speed bumps can vary considerably across exchanges (see Table \ref{tab:reforms}). In September 2015, Toronto-based exchange TSX Alpha implemented a \emph{randomized} 1 to 3 millisecond delay, to be implemented on marketable orders alone. In August 2018, Cboe (the third largest U.S. equity market operator) sought approval from the Securities and Exchange Commission (SEC) to introduce a random 3 to 4 millisecond delay on liquidity-taking, in an attempt to eliminate latency arbitrage between New York- and Chicago-based venues.\footnote{See Wall Street Journal \href{https://www.wsj.com/articles/new-speed-bump-planned-for-u-s-stock-market-1535713321}{New `Speed Bump' Planned for U.S. Stock Market}, August 31, 2018 and \href{https://www.wsj.com/articles/more-exchanges-add-speed-bumps-defying-high-frequency-traders-11564401611}{More Exchanges Add `Speed Bumps,' Defying High-Frequency Traders}, July 29, 2019.} 

\begin{table}[H]
    \caption{\label{tab:reforms} \textbf{Speed bump implementations and existing proposals}}
    
    \begin{minipage}[t]{1\columnwidth}%
\footnotesize
The table is replicated from Table 3 in \citet{Baldauf2019High-frequencyPerformance}. Randomized speed bumps are highlighted.
\end{minipage}
    \vspace{0.1in}
    \centering
    \begin{tabular}{lllll}
    \toprule
    Exchange & Country & Date & Who is delayed? & Delay  \\
    \cmidrule{1-5}
    Investors Exchange (IEX)  & United States & October 2013 & Everyone & 350 $\mu\text{s}$ \\
    \rowcolor{LightCyan} Aequitas NEO Exchange & Canada & March 2015 & Liquidity takers & 3-9 ms \\
    \rowcolor{LightCyan} TSX Alpha Exchange & Canada & September 2015 & Liquidity takers & 1-3 ms \\
    \rowcolor{LightCyan} Thomson Reuters & United States & June 2016 & All but cancel orders & 0-3 ms \\
    NYSE American & United States & July 2017 & Everyone & 350 $\mu\text{s}$ \\
    \rowcolor{LightCyan} Eurex Exchange & Germany & June 2019 & Liquidity takers & 1 or 3 ms \\
    ICE Futures & United States & Proposed & Liquidity takers & 3 ms \\
    Chicago Stock Exchange & United States & Proposed & Liquidity takers & 350 $\mu\text{s}$ \\
    NASDAQ OMX PHLX & United States & Proposed & Liquidity takers & 5 ms \\
    \rowcolor{LightCyan} Interactive Brokers & United States & Proposed & Liquidity takers & 10-200 ms \\
    \bottomrule
    \end{tabular}
\end{table}

At the other end of the spectrum, the Investors' Exchange (IEX) and NYSE American implemented in 2016 and 2017, respectively, a constant 350 microsecond delay on all orders, irrespective of whether they provide liquidity to, or consume liquidity from the market. Most recently, in May 2019, the Intercontinental Exchange (second-largest U.S. futures operator) introduced delay on trades of two precious metals contracts. At the same time, Eurex in Germany launched a pilot in June 2019 to slow down trading in options on French and German stocks.\footnote{See Financial Times, \href{https://www.ft.com/content/d99eaf40-7dfc-11e9-81d2-f785092ab560}{Futures exchanges eye shift to `Flash Boys' speed bumps}, May 29, 2019.}

Indeed, nanosecond-latency trading might not always improve market quality. \citet{MenkveldZoican2017} argue that faster exchanges promote zero-sum ``duels'' between informed HFTs and can lead to lower liquidity. \citet{biais2015equilibrium} argue that the ``arms' race'' for ever-faster markets leads to socially excessive investment in trading speed technology. Indeed, recent empirical evidence suggests that benefits from faster markets flattened out. For example, \citet{Ye2013} find that a drop in exchange latency on NASDAQ from microseconds to nanoseconds reduced market depth. \citet{Shkilko2019} find that when rain disrupts the microwave network connection between Chicago and New York, slowing down the market, liquidity actually improves. Regulators are well-aware of fast trading externalities: In 2014, the Securities and Exchange Commission (SEC) chair Mary Jo White asked ``whether the increasingly expensive search for speed has passed the point of diminishing returns.''\footnote{See the SEC Chair speech from June 5, 2014: \href{https://www.sec.gov/news/speech/2014-spch060514mjw}{Enhancing Our Equity Market Structure}.}

This papers' contribution is to study how the design of speed bumps (random or fixed, symmetric or asymmetric) impacts traders' investment in low-latency trading technology. Do trading delays help curb the speed arms' race and, if so, to what extent? Alternatively, do traders ``double down'' on low-latency technology in the presence of order delays, in an attempt to substitute exchange speed with individual speed?

Beyond speed bumps, several recent market design proposals aim to curb the negative externalities of low-latency trading. \citet{Budish2015} and \citet{HaasKhapkoZoican2018} argue that fragmenting continuous-time trading into frequent batch auctions could stimulate competition between high-frequency traders and improve liquidity. At the opposite end of the spectrum, \citet{KyleLee2017} argue for a fully-continuous exchange where traders submit buy or sell flows over time, effectively delegating order-splitting strategies to the exchange.  However, such proposal are necessarily more involved and would imply a thorough overhaul of market infrastructure and trading systems. Speed bumps are a very marginal and easy to implement change to market design, leading to wide industry adoption.

The evidence on the impact of exchange speed bumps is, so far, mixed. Studying the TSX Alpha randomized speed bumps, \citet{ChenFoley2017} document a negative impact on liquidity, while \citet{AndersonWalton2018} do not find a direct effect of the speed bump implementation on the bid-ask spread.

Empirically identifying the impact of speed bumps on the intensity of the low latency arms' race is challenging. First, investments in speed are not easily observable and could be triggered by other factors such as the emergence of new technology. Second, real-life implementation of speed bumps is often complicated by confounding effects. For example, the TSX Alpha speed bump was introduced alongside a fee structure change. Third, exchanges adopt speed bumps to obtain a competitive advantage \citep{BrolleyCimon2018}: they are consequently designed to maximize exchange profits rather than eliminate the high-frequency arms' race. In light of these challenges, we design a laboratory experiment to test the impact of speed bumps on latency investments. The experimental platform is implemented in \href{https://www.otree.org/}{oTree}, as developed in \citet{ChenSchonger2016}.

We recruited 56 management undergraduate students to participate in the trading experiment at the University of Toronto in September 2019. Participants are allocated to groups of three speculators who trade for 32 rounds on a continuous limit order book market with time priority. In each round, a profitable arbitrage opportunity opens up. Speculators submit orders to realize the arbitrage profit, but their orders arrive at the market with a delay. Time priority ensures that only the first speculator to reach the market earns the entire arbitrage profit. Moreover, a computer market maker can ``cancel'' the arbitrage opportunity, in which case all participants earn a profit of zero. The order delay has two components: an endogenous individual delay and an exogenous, common speed bump. Each round, before trading starts, speculators can invest in low-latency trading technology to reduce the expected individual delay.

Different designs of the speed bump yield different equilibrium predictions. A deterministic, symmetric speed bump (such as the one implemented in 2016 on IEX), shifts market activity uniformly in the future: As a consequence, the optimal investment in speed does not depend on the bump level. Notably, in this case, speed bumps preserve time priority: the first trader at the market wins the trade. If all snipers in the race are slowed down in the same way, their marginal valuation of speed does not change. A deterministic but asymmetric speed bump, where market makers can cancel orders without delay, leads to optimally lower investments in trading speed as speculators are at a relative disadvantage. However, an asymmetric and \emph{random} speed bump -- as recently implemented by TSX Alpha or NYSE Arca, has the potential to stimulate investment in low-latency technology and amplify the arms' race. The rationale is that speculators have stronger incentives to be fast if there is a possibility that their competitors' orders are relatively more delayed than their own.


We find empirical support for the first two of the three predictions. Without a speed bump, speculators invest 75\% of their endowment in low-latency technology (as opposed to keeping it in cash). Introducing a speed bump reduces investment, unconditionally, to 66 percent of the endowment (i.e., a 12\% decrease). The design of the speed bump is crucial: Asymmetric speed bumps reduce low-latency investment by 20\%, whereas symmetric speed bumps have little impact on traders' investment decisions. It is noteworthy that simply introducing a speed bump does not fully mitigate the arms' race, that is, reduce investment to zero. The rationale is that arbitrageurs do not only try to ``outrun'' the market-maker to the exchange, but also each other.

A second important finding is that the size of speed bump matters: increasing the speed bump magnitude by one standard deviation (for an asymmetric, deterministic delay) reduces investment in low latency by a further 8.33\%.\footnote{One standard deviation of the speed bump, in our experiment, corresponds to 2 seconds, or 40\% of the unconditional exchange latency.} The rationale is that longer speed bumps reduce the probability of \emph{any} speculator to win the race, de-emphasizing competition between speculators and enhancing the advantage of the market-maker. In contrast to \citet{Aoyagi2019SpeedBumps}, and to our simple model's prediction, we do not find any significant difference between random and deterministic speed bumps in our experimental data. 

Our results suggest that, if regulators aim to reduce investment in low-latency technology \citep[which can be detrimental to welfare, see][]{biais2015equilibrium}, the optimal policy is to implement a relatively long, asymmetric, speed bump that affects liquidity takers but not market-makers.

The remainder of the manuscript is organized as follows. Section \ref{sec:Literature} discusses the relevant literature on high-frequency trading and experimental economics. Section \ref{sec:design} presents the design of the experimental platform. Section \ref{sec:hypo} develops the empirical hypotheses that our setup can test. Section \ref{sec:results} presents the preliminary results from the laboratory experiment conducted at the University of Toronto. 

\section{Related literature \label{sec:Literature}}

Our paper builds upon an extensive literature studying the impact of trading speed and high-frequency traders (HFT) on market liquidity. \citet{Budish2015} argues that the HFT  ``arms' race'' can hurt market liquidity. \citet{MenkveldZoican2017} show that ever faster exchanges increase the frequency of zero-sum ``duels'' between fast traders, leading to lower liquidity. On the normative side, \citet{biais2015equilibrium} find that equilibrium investment in fast trading technology exceeds the social optimum. \citet{FoucaultHombertRosu2015} show that with a speed advantage, the informed investor's order flow is much more volatile, accounts for a bigger fraction of the total trading volume, and is able to forecast short-term price changes.  Empirical studies support the theoretical predictions. \citet{baron2012trading} and \citet{brogaard2014high} find HFT market orders to have a larger price impact. \citet{Shkilko2019} find that when rain disrupts microwave network and slows down trading between Chicago and New York, liquidity improves.

Several papers study ``speed bumps'' directly. In line with our prediction, \citet{Aoyagi2019SpeedBumps} argues that a random delay can incentivize investment in trading speed and worsen adverse selection. \citet{Aldrich2018OrderMessaging} study a market design featuring uniform speed bumps and find that messaging delay reduce transaction costs, but typically lead to larger queuing costs. \citet{Baldauf2019High-frequencyPerformance} find that speed bumps not only reduce transaction costs, but stimulate information production by restricting short-term speculation. \citet{BrolleyCimon2018} analyze how speed bumps impact exchange competition. They argue that the delayed exchange has better liquidity and aggregate volume increases. However, if the delayed and non-delayed exchanges belong to the same group, there are lower incentives for information acquisition and volume drops.

Empirically, \citet{ChenFoley2017} examine the market-wide effects of the introduction of a speed bump by a Canadian exchange, TSX Alpha and find lower liquidity following the reform. The TSX Alpha speed bump was asymmetric in the sense that only aggressive orders are delayed. Therefore, high-frequency liquidity providers were able to observe order flow on competing markets and cancel orders on Alpha before potential market takers could react, generating liquidity risk. \citet{AndersonWalton2018} study the same event, and find that the speed bump had no impact on market-wide liquidity. However, the authors document that frequent users of TSX Alpha experience higher fill rates and larger execution sizes, relative to less frequent users. Frequent users of TSX Alpha on average submit larger market orders and have lower order-to-trade ratios compared to less frequent users. \citet{Hu2019IntentionalExchange} finds that the SEC approval of the IEX, a speed bump pioneer, as a national securities
exchange, led to lower adverse selection risk and tighter spreads.

Finally, our paper contributes to a growing literature on experimental market microstructure. \citet{Aldrich2019} conduct a market design study on high-frequency trading and find that frequent batch auctions improve liquidity relative to continuous-time markets. \citet{bloomfield2009noise} study noise trading in an experimental setting. They find that traders with no information advantage or exogenous reason to trade behave as contrarian noise traders. While they make losses on average, they improve market liquidity by reducing bid-ask spreads and price impact, and increasing order book depth. \citet{bloomfield2015hidden} use a laboratory market experiment to investigate how the ability to hide orders affects traders' strategies and market outcomes. They find that opaque markets favor informed traders, but only when their information is particularly valuable. However, liquidity and price efficiency are not affected by opacity regimes. \citet{gozluklu2016pre} finds that market opacity enhances liquidity, especially toward the end of a trading session, and is beneficial for liquidity traders.

\section{Experimental design\label{sec:design}}

A single risky security is traded on a laboratory market with time priority. Participants submit trade orders which arrive at the market at random times, drawn from exponential distributions with endogenous rate. That is, order arrival follows a Poisson process. Further, the market may feature a ``speed bump,'' that is a further delay on participant orders. Trader profits and costs are denominated in laboratory experimental coins (ECoins), an artificial currency, and converted into Canadian dollars at the end of the experiment. 

At the start of the experiment, participants are randomly allocated to groups of three speculators, who trade against a computer-simulated market maker. In particular, each trade takes place between a participant and the computer, rather than between two participants. Once matched before the start of the first round, traders remain in the same group throughout the entire experiment. The experiment last for 32 rounds, across which we vary several market design variables:
\begin{enumerate}
    \item[(i)] the length of the speed bump;
    \item[(ii)] whether the speed bump is deterministic or random;
    \item[(iii)] whether the speed bump affects the market maker (i.e., is symmetric) or not, and
    \item[(iv)] the endowment to invest in low latency trading.
\end{enumerate}

Each round begins with an \emph{investment stage}, in which traders are given an endowment to invest in low-latency trading technology. A higher investment in low-latency increases the Poisson order arrival rate, speeding up trade execution. After each trader chooses their latency investment level, we simulate a continuous-time market where orders execute at random times against a simulated market maker. In some sense, this setup echoes the mechanics of algorithmic trading \citep[as in][]{Aldrich2019}, as traders first define strategies which are subsequently implemented by computer software. 

\paragraph{Arbitrage and time priority.} At the beginning of each round, each participant faces a profitable trading opportunity worth 100 ECoins. In the spirit of \citet{MenkveldZoican2017}, we think of the experimental traders as short-term speculators who identify an arbitrage opportunity in the form of a stale quote. 

For example, if a security is traded on two or more exchanges, it is likely that new information gets revealed on one exchange first \citep{Hagstromer2019InformationMarkets}. As a result, quotes can temporarily become misaligned (``crossed''), such that a trader can make an arbitrage profit from buying the security on one exchange at the ask price and selling it on the other at the bid. Arbitrage opportunities of this type last typically only for a few microseconds \citep{Budish2015} and are resolved either through a trade or a quote update by the market maker.\footnote{The same reasoning applies for arbitrage opportunities in quasi-identical securities traded on the same exchange, such as funds tracking the same index.}

Time priority rules makes trading speed valuable. The first participant whose order reaches the market captures the entire arbitrage profit. Any subsequent speculator orders to arrive at the market ``bounce'' against an empty limit order book and fail to execute. At the same time, the computer market maker also rushes to cancel the stale quote: if the cancel order arrives before the first participant order, all traders earn zero profit. Therefore, traders compete not only to be faster than each other, but also faster than the market maker. Formally, let $t_i$ and $t_M$ be the (stochastic) order arrival times for trader $i$ and the market maker. Trader $i$'s profit is:
\begin{equation}
    \Pi_i = \begin{cases}
            100, & \text{ if } t_i<\min\left\{t_{-i},t_M\right\}  \text{ and}\\
            0, & \text{ else.}
    \end{cases}
\end{equation}

\begin{figure}[H]
\caption{\label{fig:lom} \textbf{Continuous market and order execution}}
\vspace{0.1in}
\begin{centering}
\includegraphics[width=\textwidth]{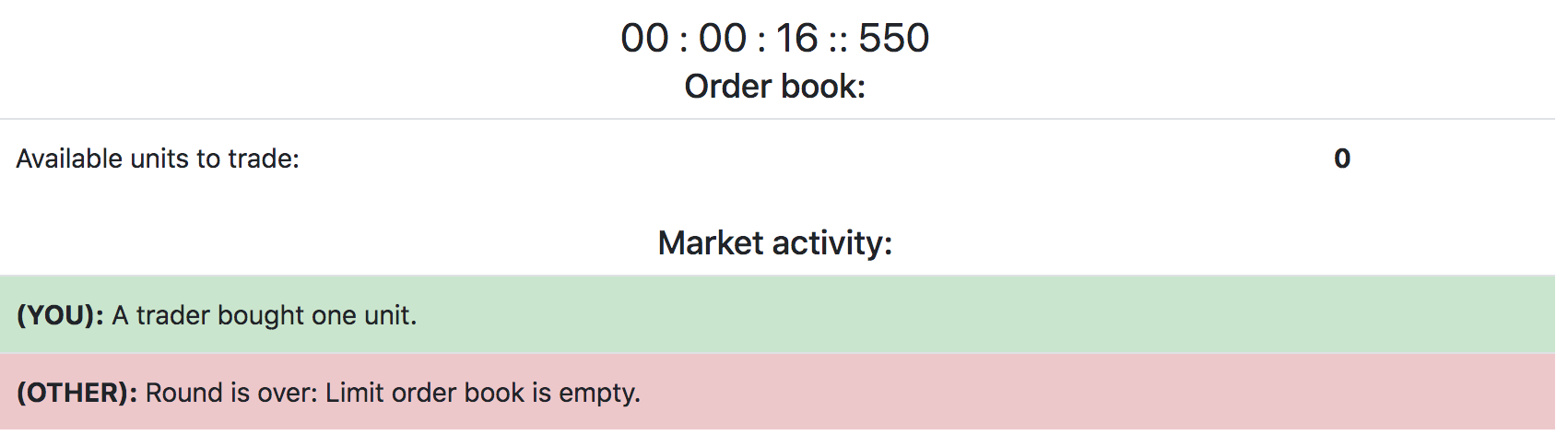}
\par\end{centering}
\end{figure}

Figure \ref{fig:lom} illustrates the limit order market screen. Participants see a trading clock (up to milisecond precision), an indicator on whether the arbitrage opportunity is still available or not, and a market activity panel with order execution.

\paragraph{Order arrival times.} For each speculator $i$, the order arrival time $t_i$ has two components: a ``trader latency'' delay $\delta_i$, drawn from an exponential distribution with parameter $\lambda_i$ and an exchange speed bump $\Delta$. The total order delay is the sum of the private, endogenous trader latency and the exogenous speed bump, 
\begin{equation}\label{eq:time_arrival}
    t_i =\delta_i \left(\lambda_i\right) + \Delta.
\end{equation}

Finally, the market maker cancels her quotes at a random time $t_M$ drawn from an exponential distribution with parameter $\lambda_M$, potentially plus the exogenous exchange speed bump.

\begin{figure}[H]
\caption{\label{fig:investment} \textbf{Investment screen}}
\begin{centering}
\includegraphics[width=\textwidth]{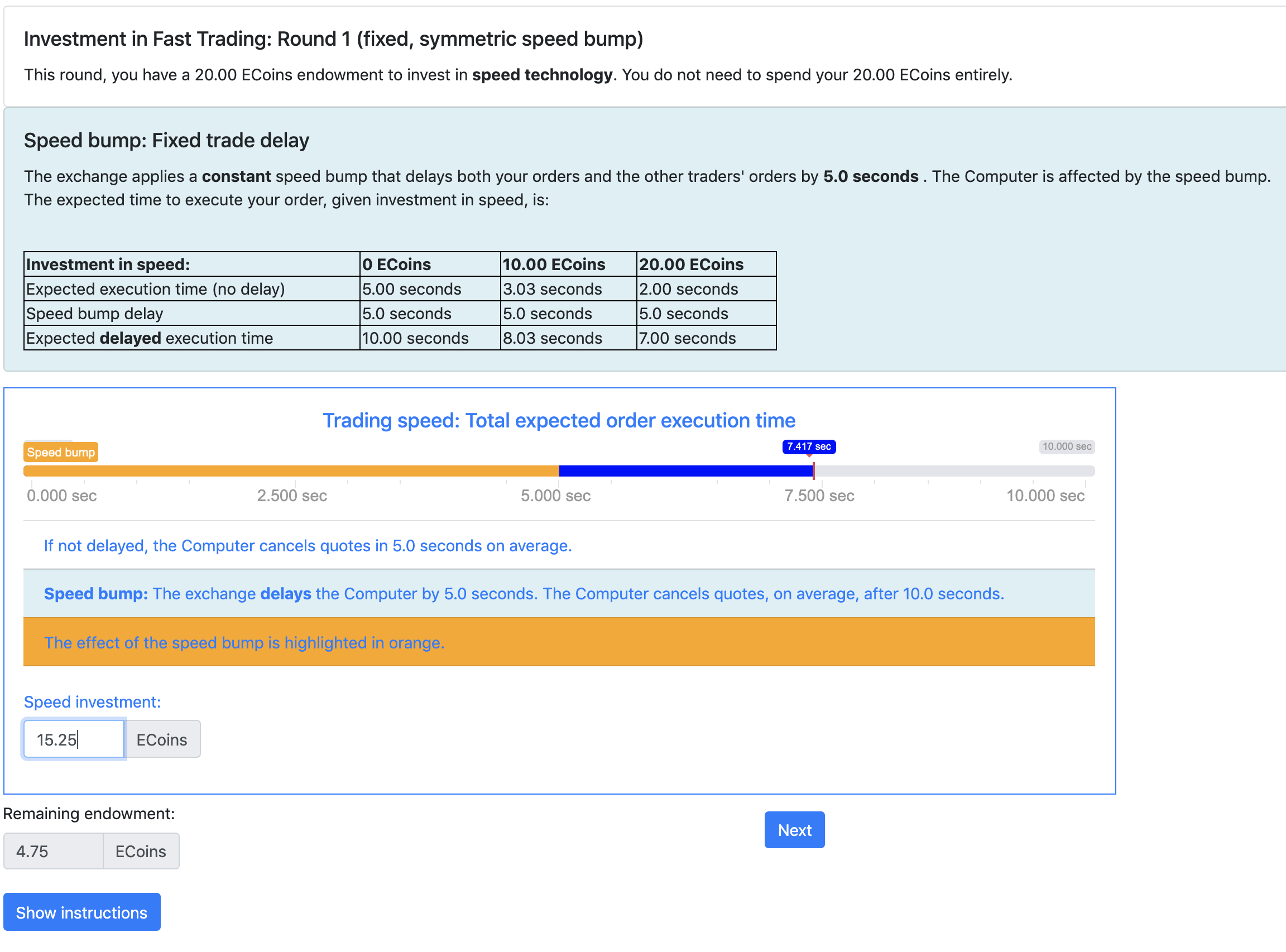}
\par\end{centering}
\end{figure}

\paragraph{Endowments.} Each round, all participants receive an endowment $\omega$ to invest in low latency technology. The endowment is the same for all participants in a given round. Participants do not need to spend the entire endowment: any portion not invested is added to their end-of-round payoff. We vary the endowment amount across rounds to be either 10 ECoins or 20 ECoins, effectively changing the opportunity cost of low latency trading in units of cash. Figure \ref{fig:investment} illustrates a typical latency investment screen for a speculator.

\subsection{Low latency technology \label{sec:tech}}

A latency investment of $\ell<\omega$ increases the order arrival rate as follows:

\begin{equation}\label{eq:latency_inv}
    \lambda = \lambda_0 + \psi\left(\frac{\ell}{\omega}\right)^\gamma, 
\end{equation}
where $\lambda_0>0$ and $\psi>0$. We set $\lambda_0=\lambda_M=0.2$, where $\lambda_M$ corresponds to the market maker's cancellation rate. On average, the computer market maker cancels quotes in 5 seconds that is, equal to $\lambda_M^{-1}$). If a participant does not invest in low latency, their expected order execution time is also 5 seconds, the reciprocal of the order arrival intensity.

The marginal impact of low latency investment on the expected arrival time is
\begin{equation}\label{eq:mr_speed}
    \frac{\partial \lambda^{-1}}{\partial \ell}=-\frac{\gamma  \psi  \left(\frac{\ell}{\omega }\right)^{\gamma }}{\ell \left(\lambda_0+\psi  \left(\frac{\ell}{\omega }\right)^{\gamma }\right)^2}<0,
\end{equation}
that is a higher investment in speed reduces the expected order arrival time. Further, a higher endowment $\omega$ reduces the marginal impact of each ECoin invested in low latency trading, since
\begin{equation}\label{eq:mr_speed_2}
    \frac{\partial^2 \lambda^{-1}}{\partial \ell \partial \omega}=\frac{\gamma ^2 \psi  \left(\frac{\ell}{\omega }\right)^{\gamma -1} \left(\lambda_0-\psi  \left(\frac{\ell}{\omega }\right)^{\gamma }\right)}{\omega ^2
   \left(\lambda_0+\psi  \left(\frac{\ell}{\omega }\right)^{\gamma }\right)^3}>0.
\end{equation}

We implement three regimes for $\left\{\psi,\gamma\right\}$, corresponding to three levels of technology productivity, or technology cost, tabulated below.
\begin{table}[H]
\caption{\label{tab:costs} \textbf{Tiers of low-latency technology (across groups)}}
    \centering
\vspace{0.1in}
\begin{center}
    \begin{tabular}{lcccc}
    \toprule
    & $\psi$ & $\gamma$ & $\lambda^{-1}\left(\ell=0\right)$ & $\lambda^{-1}\left(\ell=\omega \right)$\\
    & & & No investment & Full investment \\
    \cmidrule{1-5}
    High-cost speed technology & 0.30 & 1.25 & 5 seconds & 2 seconds \\
    Medium-cost speed technology & 0.60 & 1.50 & 5 seconds & 1.25 seconds \\ 
    Low-cost speed technology & 1.80 & 1.80  & 5 seconds & 0.5 seconds \\
    \bottomrule
    \end{tabular}
\end{center}
\end{table}

\begin{figure}
\caption{\label{fig:orderarrival} \textbf{Expected order arrival times}}
\begin{minipage}[t]{1\columnwidth}%
\footnotesize
This figure plots the expected order arrival time as a function of investment in low-latency technology (normalized by the endowment), for three trading technologies. The high-cost speed technology has parameters $\left(\psi=0.3,\gamma=1.25\right)$, whereas the medium- and low-cost speed technologies have parameters $\left(\psi=0.6,\gamma=1.5\right)$ and $\left(\psi=1.8,\gamma=1.8\right)$, respectively.
\end{minipage}
\begin{centering}

\includegraphics[width=0.85\textwidth]{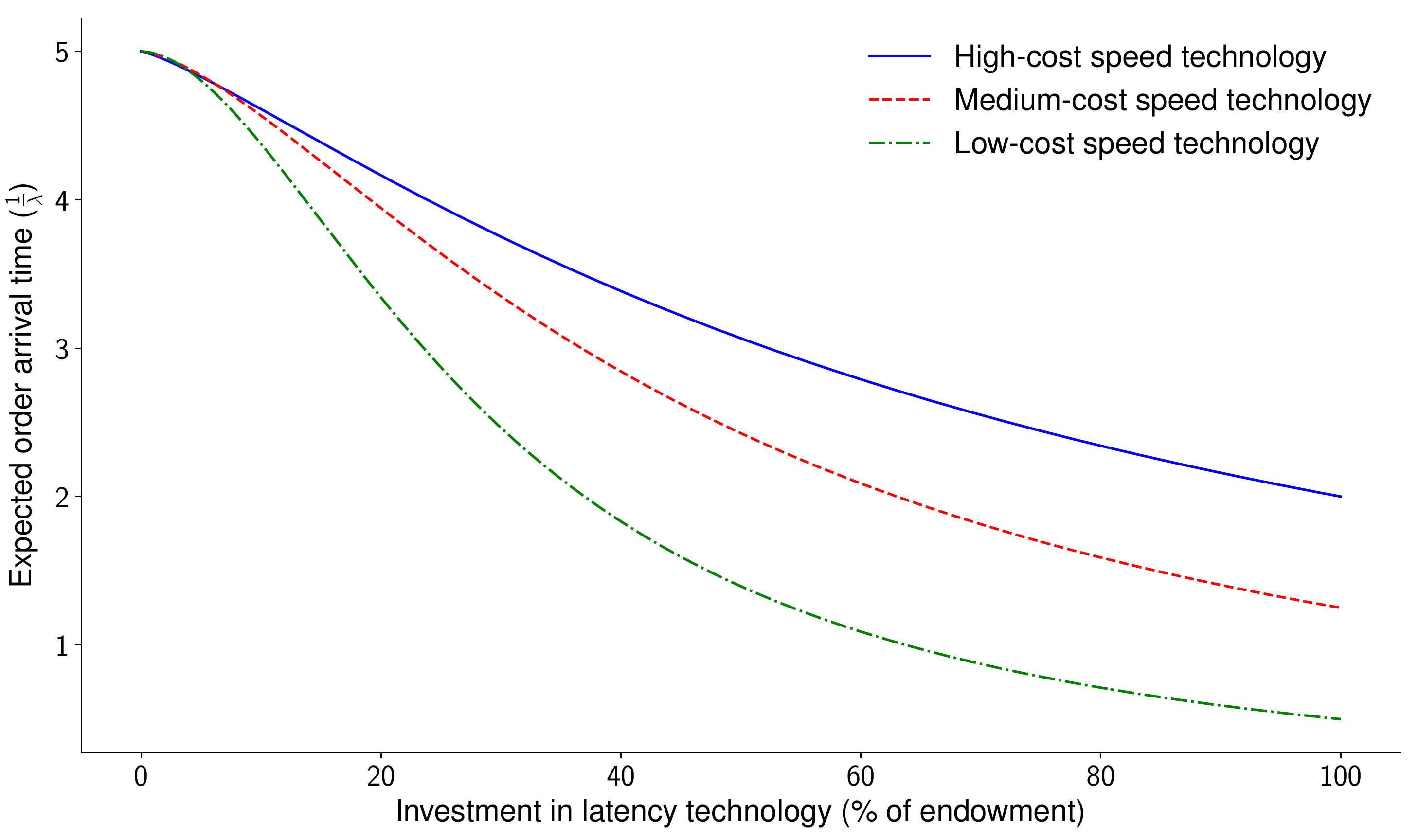}

\par\end{centering}
\end{figure}

The high-cost technology allows traders to reduce the expected arrival time from 5 to only 2 seconds conditional on investing their entire endowment. The medium- and low-cost technology allow traders to reduce their order execution time further, up to 1.25 seconds and 0.5 seconds respectively. Each group is randomly assigned one of the three technologies for the whole duration of the experiment.

\subsection{Speed bumps} 

From equation \eqref{eq:time_arrival}, each traders' order arrival time depends both on her investment in low-latency technology, as well as on an exogenous market-wide speed bump $\Delta$. The speed bump delays orders irrespective of speculators' investment in latency technology.

We consider four separate market designs for the speed bump. First, we allow the speed bump $\Delta$ to be random (for example, as implemented by the Toronto Stock Exchange in 2015) or deterministic (as implemented by the IEX in 2016). Second, we distinguish between symmetric speed bumps that affect both human traders and the computer market maker, and asymmetric speed bumps that affect human traders but not the market maker. 

\paragraph{Deterministic / random speed bumps.} We implement three deterministic delays, of one, three, and five seconds respectively. The deterministic delays are of the same order of magnitude with the traders' own latency (which is capped at 5 seconds, as discussed in Section \ref{sec:tech}).  

We implement random speed bumps as follows. For a given average speed bump size $\Delta$, we independently draw for each trader one of three equally likely delays : either $0.5\times\Delta$ (low delay), $\Delta$, or $1.5\times \Delta$ (high delay).

\begin{center}
    \begin{tabular}{ccccc}
    \toprule
    & & \multicolumn{3}{c}{Speed bump size} \\
    & No speed bump & Small & Medium & Large \\
    \cmidrule{1-5}
    $\underline{\Delta}=0.5\Delta$ & 0 & 0.5 seconds & 1.5 seconds & 2.5 seconds \\
    $\Delta$ & 0 & 1 second & 3 seconds & 5 seconds \\
    $\overline{\Delta}=1.5\Delta$ & 0 & 1.5 seconds & 4.5 seconds & 7.5 seconds \\ 
    \bottomrule
    \end{tabular}
\end{center}

The implementation is consistent with random speed bumps as implemented by exchanges such as TSX Alpha or Cboe, who add a probabilistic delay to orders \citep{BrolleyCimon2018}. 

\paragraph{Symmetric / asymmetric speed bumps.} Further, we distinguish between symmetric speed bumps, where both human traders and the computer market maker orders are delayed, and asymmetric speed bumps, where only human traders are delayed.

The expected order arrival time for a speculator is
\begin{equation}\label{eq:expected_time}
    \mathbb{E} t_i = \frac{1}{\lambda_i} + \Delta,
\end{equation}
which is always larger than 1 second (0.5 seconds from trader latency with full investment under low-cost technology, plus a 0.5 seconds speed bump) and always smaller than 12.5 seconds (5 seconds from trader latency with zero investment plus a 7.5 seconds speed bump).

\subsection{Trading rounds}

The experiment lasts for 32 rounds. The first 4 rounds are training rounds and are discarded in the data analysis. Out of the following 28 rounds, 24 rounds combine three speed bump sizes (small/medium/large) with four speed bump design choices and two endowment levels, that is $3 \times 4 \times 2 =24$. Finally, four rounds have no speed bump. The sequence of rounds is randomized as in Table \ref{tab:rounds}, to reduce the impact of learning over time.

\begin{table}[H]
\caption{\label{tab:rounds} \textbf{Trading round sequence}}
\begin{center}
    \begin{tabular}{ccccc|ccccc}
    \toprule
    \# & $\Delta$ & Symmetric? & Random? & Endowment & \# & $\Delta$ & Symmetric? & Random? & Endowment \\
    \cmidrule{1-10}
1 & 	5 & 	Yes & 	No &  20 & 	17 & 		3 & 	Yes & 	Yes & 	20 \\ 
2 & 	1 & 	No & 	Yes & 	10 &  18 & 		5 & 	No & 	No & 	10 \\ 
3 & 	3 & 	Yes & 	Yes & 	20 &  19 & 		1 & 	No & 	Yes & 	20 \\ 
4 & 	0 & 	No & 	No & 	10 & 20 & 	 	5 & 	Yes & 	No & 	20 \\ 
\cmidrule{1-5}
5 & 	1 & 	Yes & 	Yes & 	10 & 21 & 	 	3 & 	No & 	No & 	10 \\ 
6 & 	5 & 	No & 	Yes &  20 & 	22 & 		0 & 	 & 	 & 	20 \\ 
7 & 	3 & 	Yes & 	Yes & 	10 & 23 & 	 	5 & 	Yes & 	Yes & 	20 \\ 
8 & 	1 & 	Yes & 	No & 		20 & 24 &  	3 & 	Yes & 	No & 	10 \\ 
9 & 	5 & 	Yes & 	Yes & 	10 & 25 & 	 	1 & 	No & 	Yes & 	10 \\ 
10 & 	0 & 	 & 	 & 	20 & 26 & 	 	3 & 	No & 	No & 	20 \\ 
11 & 	1 & 	Yes & 	No & 	10 & 27 & 	 	5 & 	No & 	Yes & 	10 \\ 
12 & 	5 & 	No & 	No & 	20 &  28 & 		0 & 	 & 	 & 	10 \\ 
13 & 	3 & 	No & 	Yes & 		10 & 29 &  	1 & 	No & 	No & 	20 \\ 
14 & 	1 & 	Yes & 	Yes & 	20 & 30 & 	 	3 & 	Yes & 	No & 	20 \\ 
15 & 	5 & 	Yes & 	No & 	10 & 31 & 	 	3 & 	No & 	Yes & 	20 \\ 
16 & 	0 & 	 & 	 & 	10 & 32 & 	 	1 & 	No & 	No & 	10 \\ 
    \bottomrule
    \end{tabular}
\end{center}
\end{table}

\begin{figure}[H]
\caption{\label{fig:payoffs_fig} \textbf{End-of-round payoff announcements}}
\vspace{0.1in}
\begin{centering}

\includegraphics[width=\textwidth]{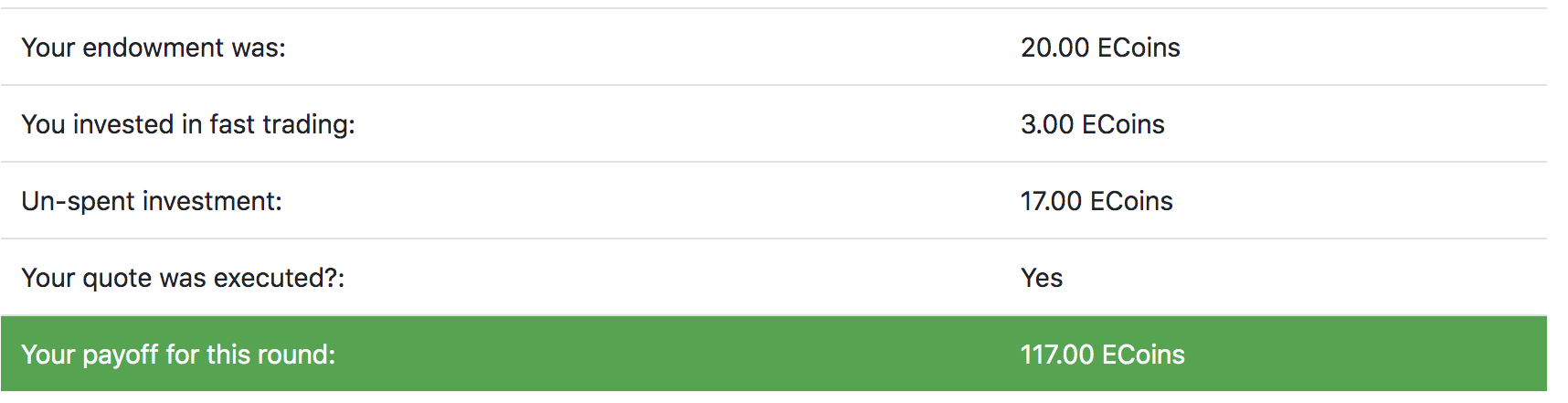}

\par\end{centering}
\end{figure}

After each round, traders see a ``Results'' screen confirming whether their order was executed and their payoff for the round, as illustrated in Figure \ref{fig:payoffs_fig}.

\paragraph{Elicitation of risk-aversion.} We elicited risk-aversion using standard \citet{HoltLaury2002} choice lists and the Bomb Risk Elicitation Task (BRET) described in \citet{Crosetto2013}. Each participant fulfills both risk-aversion tasks in a random order. Finally, all participants are required to fill in a demographic questionnaire. The flowchart of the experiment is detailed below:

\begin{center}
\includegraphics[width=0.9\textwidth]{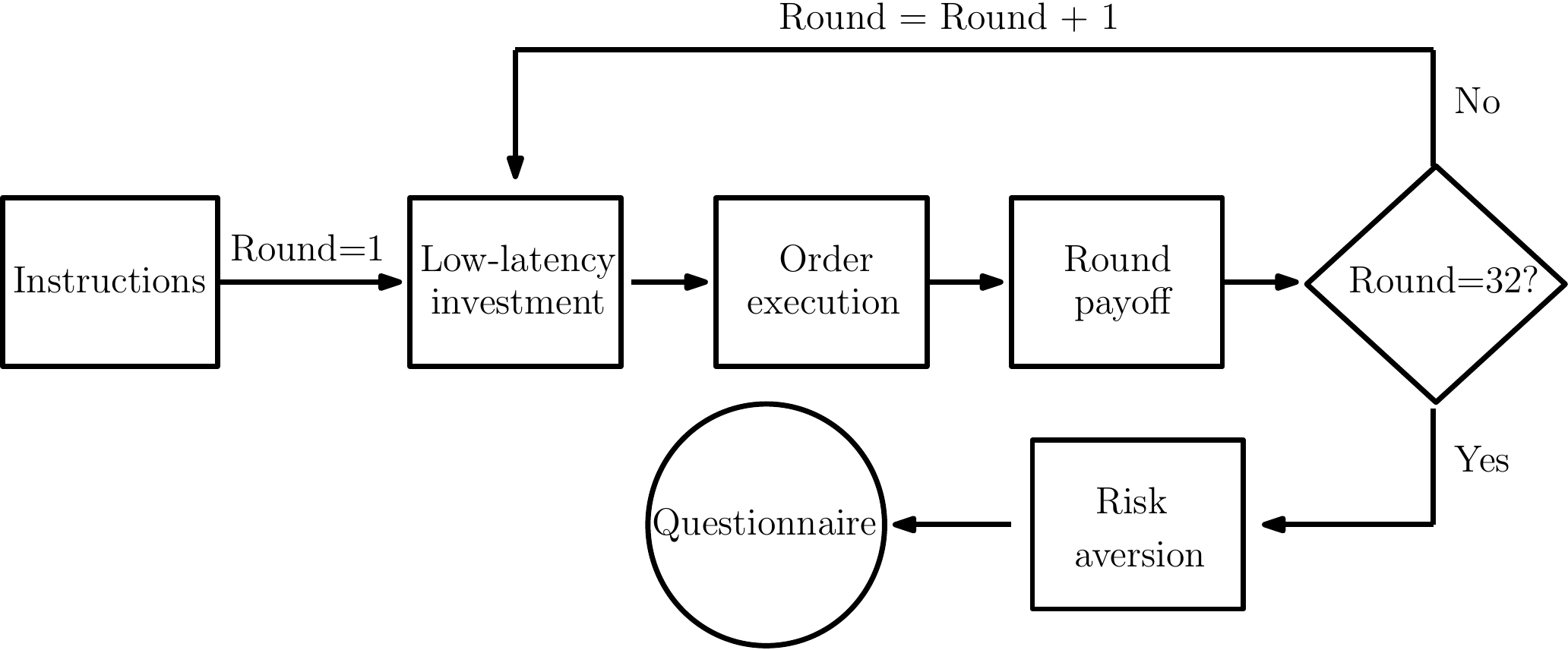}  
\end{center}

\begin{figure}[H]
\caption{\label{fig:instructions} \textbf{On-screen instructions at the beginning of the experiment}}
\vspace{0.1in}
\begin{centering}
\includegraphics[width=\textwidth]{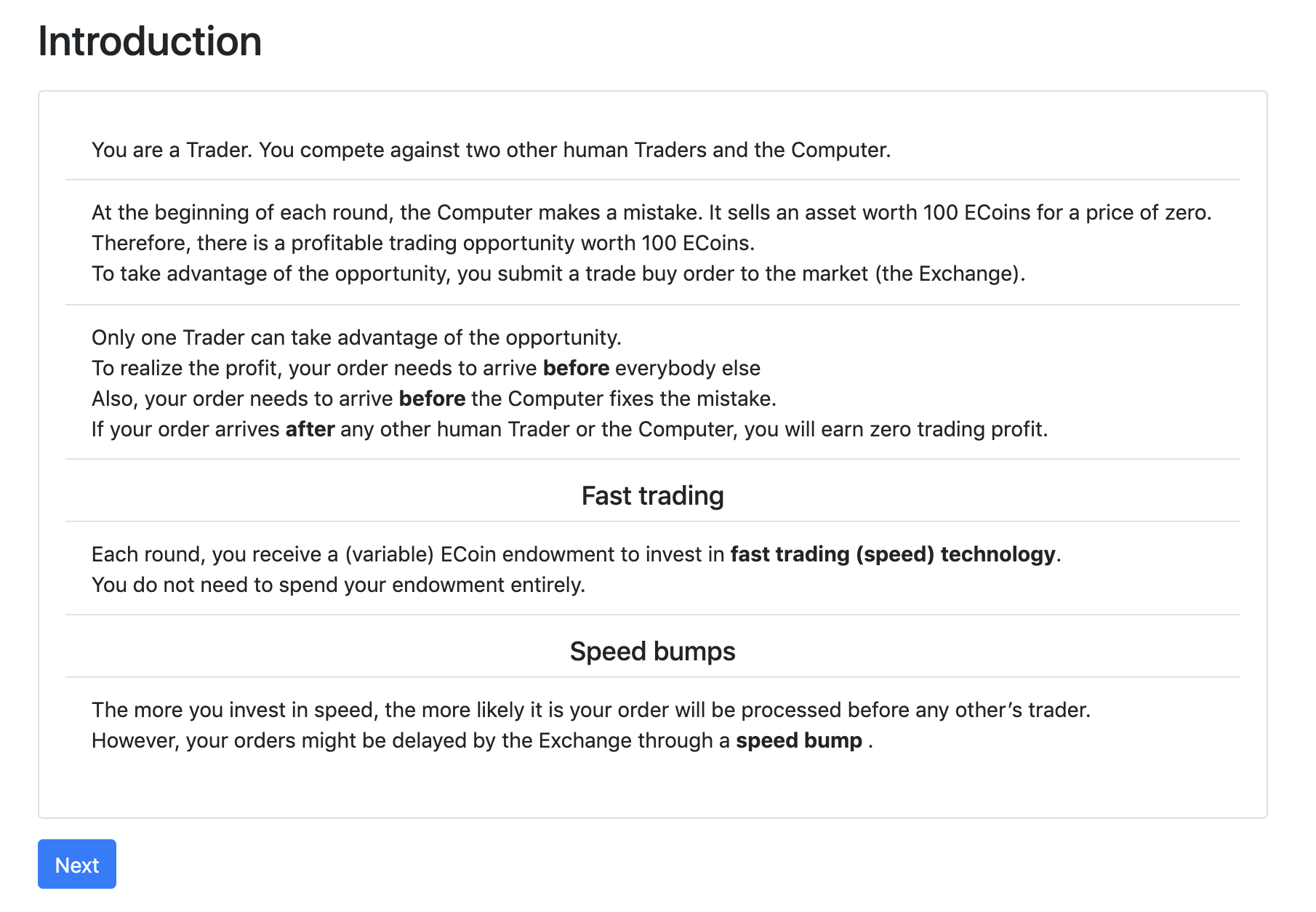}
\par\end{centering}
\end{figure}

Figure \ref{fig:instructions} displays the instructions shown to each trader at the start of the experiment.

\section{Hypotheses development \label{sec:hypo}}

In this Section we elaborate testable hypotheses that follow from the experiment design discussed in Section \ref{sec:design}. For simplicity of exposition, we illustrate the economic mechanisms in a simplified setup with two traders and one market maker. We write the expected profit of trader $i$ as
\begin{equation}\label{eq:profit_hyp}
    \Pi_i = \text{Prob}\left(i \text{ trade is executed} \mid \lambda_i \right)\times \Sigma - \mathcal{C}\left(\lambda_i\right), 
\end{equation}
where $\Sigma=100 \text{ ECoins}$, that is the size of the profit opportunity, and $\mathcal{C}\left(\cdot\right)$ is the cost of investing in low-latency technology.

We focus first on deterministic, but potentially asymmetric, speed bumps. In particular, human traders/speculators face a speed bump $\Delta$ and the market maker faces a speed bump $\Delta_M\in\left\{0,\Delta\right\}$. In this case, trader $i$'s order is executed if and only if it arrives first at the market:
\begin{equation}
    t_i+\Delta \leq  \min \left\{t_{-i}+\Delta, t_M+\Delta_M\right\},
\end{equation}
where $t_i$ is the stochastic order arrival time for trader $i$ without the speed bump. From the definition of the exponential distribution, it follows that the execution probability for trader $i$ is
\begin{equation}\label{eq:prob_det}
    \text{Prob}\left(t_i\leq \min \left\{t_{-i},t_M+\left(\Delta_M-\Delta\right)\right\}\right)=e^{\lambda_M\left(\Delta_M-\Delta\right)}\frac{\lambda_i}{\lambda_i+\lambda_{-i}+\lambda_M}.
\end{equation}

\begin{hypth}\label{h:h1}
\begin{leftbar} \setlength{\parskip}{0ex}
If the speed bump is \textbf{symmetric} and \textbf{deterministic}, investment in low-latency technology does not depend on the speed bump magnitude.
\end{leftbar}
\end{hypth}

Hypothesis \ref{h:h1} is straightforward. A symmetric and deterministic speed bump delays all orders by the same fixed amount, including the market maker's cancellation message. In particular, the speed bump simply therefore shifts all market activity into the future, and does not alter the order in which messages are processed by the exchange. From equation \eqref{eq:prob_det}, if $\Delta_M=\Delta$, then the execution probability for trader $i$ does not depend on the speed bump level. Therefore, the optimal latency investment corresponds to the arrival rate $\lambda^\star_i$ which solves the first order condition
\begin{equation}\label{eq:foc}
    \Sigma \times \frac{\partial \text{Prob}\left(i \text{ trade is executed} \mid \lambda_i \right)}{\partial \lambda_i} \left(\lambda^\star_i\right) = \mathcal{C}^\prime\left(\lambda^\star_i\right),
\end{equation}
where neither the left- or the right-hand side of \eqref{eq:foc} depend on the level of the speed bump.

\begin{hypth}\label{h:h2}
\begin{leftbar} \setlength{\parskip}{0ex}
If the speed bump is \textbf{asymmetric} and \textbf{deterministic}, investment in low-latency technology decreases in the speed bump magnitude.
\end{leftbar}
\end{hypth}

An asymmetric speed bump implies $\Delta_M=0$: that is, the market maker cancellation order is not delayed. From equation \eqref{eq:prob_det}, the marginal benefit of investment in low latency decreases in the speed bump magnitude, since if $\Delta_M=0$ it follows that
\begin{equation}\label{eq:partial_derivative}
    \frac{\partial \text{Prob}\left(i \text{ trade is executed} \mid \lambda_i \right)}{\partial \lambda_i} = e^{-\lambda_M \Delta}\frac{\lambda_j+\lambda_M}{\left(\lambda_i+\lambda_{-i}+\lambda_M\right)^2}.
\end{equation}
Therefore, a deterministic, asymmetric speed bump reduces the incremental benefit of low-latency technology and does not affect the marginal cost. It follows from the first-order condition \eqref{eq:foc} that the optimal investment in low-latency technology decreases in the magnitude of the speed bump.

\begin{hypth}\label{h:h3}
\begin{leftbar} \setlength{\parskip}{0ex}
If the speed bump is \textbf{symmetric} and \textbf{random}, investment in low-latency technology increases in the speed bump magnitude.
\end{leftbar}
\end{hypth}

If the speed bump is symmetric and random, traders have stronger incentives to invest in speed. From equations \eqref{eq:prob_det} and \eqref{eq:foc}, the execution probability is convex in the size of the speed bump. Since the speed bump is random, trader $i$'s competitors can be more or less delayed than $i$. From equation \eqref{eq:partial_derivative}, traders optimally increase investment in speed when competitors face a relatively longer delay and decrease investment in speed otherwise. Convexity ensures that the first effect dominates. In some sense, traders behave as they have a preference for risk. 

In the experiment implementation, there are three possible speed bump values: that is, $\tilde{\Delta}_i\in\left\{\frac{1}{2}\Delta,\Delta,\frac{3}{2}\Delta\right\}$. Therefore, there are $3^3=27$ equally likely states of the world to characterize the relative delay of $i$ compared to $-i$ and to the market maker, respectively. Formally, we note that the function
\begin{equation}
    f\left(x, \Delta\right)=e^{\Delta x} + e^{-\Delta x}
\end{equation}
increases in $\Delta$.\footnote{The function $f\left(x\right)$ is a linear transformation of the standard hyperbolic cosine function, which increases for $x\geq 0$.} It can be shown that the execution probability for trader $i$ can be written as a sum of functions increasing in $\Delta$, that is
\begin{equation}\label{eq:prob_exec_random}
\text{Prob}\left(i \text{ trade is executed} \mid \lambda_i, \tilde{\Delta} \right)=\sum_k \phi_{0,k} f\left(\phi_{1,k} \lambda_{-i} + \phi_{2,k} \lambda_M, \Delta\right)  \times \frac{\lambda_i}{\lambda_i+\lambda_{-i}+\lambda_M},
\end{equation}
where the coefficients $\phi_{0,k}>0$ and $\phi_{1,k}, \phi_{2,k} \in \left\{0,\pm \dfrac{1}{2}, \pm 1\right\}$.\footnote{The closed-form expression for the sum in \eqref{eq:prob_exec_random} is $\frac{1}{27} \Psi$, where
\begin{align}
    \Psi&= 3+f\left(\lambda_{-i}\right)+f\left(\lambda_M\right)+f\left(\lambda_{-i}+\lambda_M\right)+f\left(0.5\lambda_{-i}+\lambda_M\right)+f\left(\lambda_{-i}+0.5\lambda_M\right) \nonumber \\ 
    &+f\left(0.5\lambda_{-i}-0.5\lambda_M\right) +2f\left(0.5\lambda_{-i}\right)+2f\left(0.5\lambda_M\right)+2f\left(0.5\lambda_{-i}+0.5\lambda_M\right). \nonumber 
\end{align}
Appendix \ref{sec:states} enumerates the relative delays in all 27 states of the world.} Therefore, the execution probability as well as the marginal benefit of speed investment increase in $\Delta$: traders optimally invest more in low-latency technology as the length of a random, symmetric speed bump increases.

A random, asymmetric speed bump yields a non-linear relationship between speed bump size and low-latency investment. On the one hand, traders have incentive to reduce speed investment as they are at a disadvantage relative to the market maker (the mechanism in Hypothesis \ref{h:h2}). On the other hand, the randomness component stimulates speed investment as speculators compete with each other (the mechanism in Hypothesis \ref{h:h3}). The main experimental predictions are summarized in Table \ref{tab:hyp}.

\begin{table}[H]
\caption{\label{tab:hyp} \textbf{Speed bump design and low-latency investment}}
    \centering
\vspace{0.1in}
\begin{tabular}{|m{3em}|c|c|}
\hline
\makecell{Speed \\ bump} & Deterministic & Random \\
\hline
\hspace{2ex}\rotatebox{90}{Asymmetric } & \makecell{Reduces speed \\ investment \\ \emph{(hypothesis \ref{h:h2})}} & \makecell{Non-linear \\ effect}\\
\hline
\hspace{2ex}\rotatebox{90}{Symmetric } & \makecell{No effect on \\ investment \\ \emph{(hypothesis \ref{h:h1})}} & \makecell{Stimulates speed \\ investment \\ \emph{(hypothesis \ref{h:h3})}}\\
\hline
\end{tabular}
\end{table}

Two additional hypotheses, discussed below, test the impact of changing the cost of low-latency on trading speed acquisition by experimental participants.

\begin{hypth}\label{h:h4}
\begin{leftbar} \setlength{\parskip}{0ex}
Investment in low-latency technology decreases in the endowment $\omega$.
\end{leftbar}
\end{hypth}

The intuition driving Hypothesis \ref{h:h4} is that from equation \eqref{eq:latency_inv}, the order arrival rate is driven by the relative investment in low latency, $\dfrac{\ell}{\omega}$, where $\omega$ is the original endowment. That is, participants need to invest 2 ECoins out of an endowment of 20 ECoins (10\%) to achieve the same order arrival rate as if they invested 1 ECoin out of a 10 ECoins endowment. Therefore, a higher endowment increases the opportunity cost of cash and should reduce investment in low latency. Hypothesis \ref{h:h5} uses the cross-sectional variation in technology cost to test whether participants reduce investment when technology becomes more expensive.

\begin{hypth}\label{h:h5}
\begin{leftbar} \setlength{\parskip}{0ex}
Participants in groups with high-cost speed technology invest less in speed than participants in groups with low-cost technology, where technology cost is defined in Table \ref{tab:costs}.
\end{leftbar}
\end{hypth}

\section{Experimental Results \label{sec:results}}

\subsection{Cohort formation}
Participants to the experiment were recruited from the University of Toronto cohort of undergraduate students, using the campus Study Pool Log-in System (Sona). The experimental session took place on Tuesday, September 24, 2019 at the Bridge Lab in the Scarborough campus, and lasted for two and a half hours. 

In total, 56 students participated in the pilot experiment. They were randomly assigned to 16 groups of 3 students and 2 groups of 4 students. Participants received 2.5 course credits for attendance, and a variable cash amount disbursed in the form of Amazon e-Gift cards. The currency rate was set to \$0.50 (Canadian) for 1 ECoin.

\paragraph{Demographics.} The participants completed a demographics questionnaire after the experiment ended. Participants are between 19 and 23 years old, with a median and modal age of 20. Gender-wise, 36 traders identify as female and 20 traders traders identify as male. Twenty-six out of the 56 students are finance majors (46.4\% of the sample), 8 students major in economics (14.28\%), and 13 students focus on other management-related majors (23.21\%). A majority of the sample (58.93\%) took at least one course in finance before the experiment. The sample composition by year of study is tabulated below. 

\begin{center}
    \begin{tabular}{ccc|ccc}
    \toprule
    Education level & Participants & Share & Finance courses & Participants & Share \\
    \cmidrule{1-6}
    Second-year undergraduate & 9 & 16.07\% & No courses & 23 & 41.07\% \\
    Third-year undergraduate  & 37 & 66.07\% & One course & 19 & 33.02\%  \\
    Fourth-year undergraduate & 10 & 17.86\% & Two courses & 9 & 16.07\% \\
    & & & Three courses & 5 & 8.92\% \\
    \bottomrule
    \end{tabular}
\end{center}

Finally, 14 out the 56 participants (25\%) who completed questionnaires reported they have prior trading experience, and 19 out of 56 (that is, 33.4\% of the sample) have participated in experiments before.

\subsection{Summary statistics}

Do speed bumps, on average, reduce investment?
Our main dependent variable is the fraction of trader's endowment directed towards investment in the low-latency technology, $\dfrac{\ell}{\omega}$. In the absence of a speed bump, traders invested 75\% of their endowment in speed, on average. If a speed bump is introduced, the average relative investment in speed drops to 66\%. However, as conjectured in Section \ref{sec:hypo}, different types of speed bumps affect traders incentives to invest in speed differently.

Table \ref{tab:averages} presents the average levels of investment in speed for different speed bumps designs. When facing symmetric speed bumps, traders did not significantly reduce their investment in speed. Asymmetric speed bumps, on the other hand, reduced investments in low-latency technology by 20.6 percent. Since traders compete with each other, and not only with the computer market maker, a speed bump does not drive investment to zero.

\begin{table}[H]
\caption{\label{tab:averages} \textbf{Average low-latency investment}}
    \centering
\vspace{0.1in}
\begin{tabular}{cccc}
\toprule
\textbf{Speed bump} & Deterministic & Random & No speed bump \\
\cmidrule{1-4}
Asymmetric  & 60.32\% & 58.72\% & \multirow{2}{*}{74.99\%} \\
Symmetric & 75.10\% & 71.07\% & \\
\bottomrule
\end{tabular}
\end{table}

Table \ref{tab:reg1} formally investigates if such differences in means are statistically significant. To this end we estimate a simple regression model with multi-level fixed effects:
\begin{equation}
    \frac{\ell_{i,j,t}}{\omega_{i,j,t}}=\psi_{0,i,j}+\psi_1 d_\text{speed bump} +\varepsilon_{i,j,t},
\end{equation}
where $i$ runs over traders, $j$ runs over groups of traders, and $t$ indexes trading rounds.

We observe that the overall reduction of speed investment in the presence of a speed bump is statistically significant and is largely driven by the rounds with asymmetric speed bumps. As conjectured in Section \ref{sec:hypo}, the introduction of a symmetric and deterministic speed bump is not statistically different from the market with no speed bump at all. In line with the intuition outlined in Section \ref{sec:hypo}, traders reduced their speed investment when facing an asymmetric speed bump which was putting them at a disadvantage relative to the market maker. This statistically and economically significant decrease in investment can be observed when asymmetric speed bumps are either deterministic or random. Finally, symmetric and random speed bumps did not significantly alter traders' choices in terms of speed investment. 

\begin{table}[H]
\caption{\label{tab:reg1} \textbf{Average low-latency investment}}
    \centering
\vspace{0.1in}
\begin{tabular}{lcc}
\toprule
Sample & No speed bump ($\psi_0$) & Impact of speed bump ($\psi_1$) \\
\cmidrule{1-3}
Overall &  0.75  & -0.09** \\
& & (-2.65) \\
\cmidrule{1-3}
Random & 0.75 & -0.10** \\
& & (-2.89) \\
Deterministic & 0.75 & -0.07 \\
& & (-1.75) \\
\cmidrule{1-3}
Symmetric & 0.75 & -0.02 \\
& & (-0.63) \\
Asymmetric & 0.75 & -0.15***\\
& & (-3.89) \\
\cmidrule{1-3}
Random and symmetric & 0.75 & -0.04 \\
& & (-1.27) \\
Random and asymmetric & 0.75 & -0.16*** \\
& & (-3.87) \\
Deterministic and symmetric & 0.75 & 0.00 \\
& & (0.03) \\
Deterministic and asymmetric & 0.75 & -0.14** \\
& & (-3.00) \\
\bottomrule
\multicolumn{3}{c}{Robust t-statistics in parentheses. *** p$<$0.01, ** p$<$0.05, * p$<$0.1.} \\
\multicolumn{3}{c}{Three-way clustered standard errors (participant, group, and period).} \\
\end{tabular}
\end{table}

\subsection{Regression analysis}

In this subsection we formally test the hypotheses postulated in Section \ref{sec:hypo}. To this end, we estimate the following panel model:

\begin{equation} \label{eq:main}
    \frac{\ell_{i,j,t}}{\omega_{i,j,t}}=\phi_{0,i,j}+\phi_1 d_s + \phi_2 d_r + \phi_3 d_s d_r + \Delta \left(\phi_4+\phi_5 d_s + \phi_6 d_r + \phi_7 d_s d_r\right)+\text{Controls}+\varepsilon_{i,j,t},
\end{equation}
where $i$ runs over traders, $j$ runs over groups of traders, and $t$ indexes trading rounds. The dummy variables $d_s$ and $d_r$ are the indicators for the symmetric and random speed bumps, respectively, and $\Delta$ is the standardized speed bump size (rescaled to have a mean of zero and a standard deviation of one). The model includes fixed effects for traders, groups of traders, and game rounds. We control for investment endowment (standardized) and the indicator for winning the previous trading round. Standard errors are clustered at trader, group, and round level. The estimation results of regression model \eqref{eq:main} are presented in Table \ref{tab:main}.

\begin{table}[H]
\caption{\label{tab:main} \textbf{Impact of speed bumps on low-latency investment}}
    \centering
\vspace{0.1in}

\begin{tabular}{lllllll} 
\toprule
& \multicolumn{6}{c}{Relative investment in speed} \\
 & (1) & (2) & (3) & (4) & (5) & (6) \\
\cmidrule{2-7}
Constant & 0.60*** & 0.60*** & 0.60*** & 0.60*** & 0.60*** & 0.60*** \\
 & (36.89) & (37.93) & (34.49) & (33.95) & (34.38) & (31.79) \\
Speed bump size ($\Delta$) & -0.05** & -0.05** & -0.05** & -0.05*** & -0.05*** & -0.05** \\
 & (-2.61) & (-2.60) & (-2.20) & (-3.10) & (-3.09) & (-2.45) \\
Symmetric speed bump ($d_\text{s}$) & 0.15*** & 0.15*** & 0.15*** & 0.15*** & 0.15*** & 0.15*** \\
 & (4.36) & (4.36) & (4.12) & (4.21) & (4.22) & (3.99) \\
Random speed bump ($d_\text{r}$) & -0.02 & -0.02 & -0.02 & -0.02 & -0.02 & -0.02 \\
 & (-0.69) & (-0.70) & (-0.67) & (-0.64) & (-0.66) & (-0.63) \\
Symmetric and random ($d_\text{s} \times d_\text{r}$) & -0.02 & -0.02 & -0.02 & -0.02 & -0.02 & -0.02 \\
 & (-1.43) & (-1.40) & (-1.17) & (-0.89) & (-0.88) & (-0.78) \\
Size $\times$ symmetric ($d_\text{s} \times \Delta$) & 0.04* & 0.04* & 0.04 & 0.04** & 0.04** & 0.04 \\
 & (1.93) & (1.92) & (1.40) & (2.12) & (2.15) & (1.45) \\
Size $\times$ random ($d_\text{r} \times \Delta$)  & 0.02 & 0.02 & 0.02 & 0.02 & 0.02 & 0.02 \\
 & (0.80) & (0.79) & (0.79) & (0.88) & (0.87) & (0.85) \\
Size $\times$ symmetric and random & -0.03 & -0.03* & -0.03 & -0.03 & -0.03 & -0.03 \\
 & (-1.70) & (-1.93) & (-1.30) & (-1.55) & (-1.52) & (-1.21) \\
Investment endowment & -0.02 & -0.02 &  & -0.02* & -0.02* &  \\
 & (-1.70) & (-1.70) &  & (-1.80) & (-1.80) &  \\
Won in previous round & 0.00 &  & 0.00 & 0.00 &  & 0.00 \\
 & (0.33) &  & (0.21) & (0.26) &  & (0.17) \\
Observations & 1,344 & 1,344 & 1,344 & 1,344 & 1,344 & 1,344 \\
R-squared & 0.22 & 0.22 & 0.22 & 0.22 & 0.22 & 0.22  \\
Participant FE & Yes & Yes & Yes & Yes & Yes & Yes \\
Group FE & Yes & Yes & Yes & No & No & No \\
\cmidrule{1-7}
Observations & 1,344 & 1,344 & 1,344 & 1,344 & 1,344 & 1,344 \\
R-squared & 0.22 & 0.22 & 0.22 & 0.22 & 0.22 & 0.22 \\
Participant FE & Yes & Yes & Yes & Yes & Yes & Yes \\
 Group FE & Yes & Yes & Yes & Yes & No & No \\
\bottomrule
\multicolumn{7}{c}{Robust t-statistics in parentheses. *** p$<$0.01, ** p$<$0.05, * p$<$0.1.} \\
\multicolumn{7}{c}{Three-way clustered standard errors (participant, group, and period).} \\
\end{tabular}
\end{table}

Table \ref{tab:main} presents the main findings of the paper. First, asymmetric speed bumps (both random and deterministic) correspond to the lowest investment in speed technology. 

We find strong support for Hypothesis \ref{h:h2}: if an asymmetric, deterministic speed bump is implemented, a longer message delay further deters investment in low-latency technology. If the speed bump magnitude increases by one standard deviation, traders reduce their investment in speed by 8.33\% (0.05/0.60, i.e., coefficient $\phi_4$ in the regression model). In our setup, a standard deviation of the speed bump corresponds to a delay of 2 seconds, equivalent to 40\% of the unconditional exchange latency with no investment. Alternatively, moving from a small speed bump (20\% of the unconditional exchange latency) to a large one (100\% of the unconditional exchange latency) further reduces investment in speed by 16.7\%.

To formally test Hypothesis \ref{h:h1}, we check whether the sum of $\phi_4$ (coefficient of $\Delta$) and $\phi_5$ (coefficient of $d_\text{s}\Delta$) is different from zero. We find that the sum of the two coefficients is -0.01 (-0.05+0.04), statistically indistinguishable from zero at a 5\% confidence level (the p-value of the corresponding F-test is 0.099).  Therefore, our experimental results support Hypothesis \ref{h:h1}: the magnitude of a symmetric and deterministic speed bump does not impact low-latency investment.

From Table \ref{tab:main}, the market design choice between a random and deterministic speed bump has little impact on traders' investment decision. That is, we do not find evidence in line with Hypothesis \ref{h:h3}, which postulates that if the speed bump is symmetric and random, traders' speed investment increases in trading delay. In order to test this hypothesis, we look at the sum of coefficients $\phi_4$, $\phi_5$, $\phi_6$, and $\phi_7$ (coefficients on $\Delta$ and all the dummy interactions with $\Delta$). This sum is not statistically different from zero (the p-value of the corresponding F-test is 0.054). 


Finally, results in Table \ref{tab:main} provide some weak evidence that speed investment decreases in traders' endowment as postulated in Hypothesis \ref{h:h4}. A larger investment endowment increases the opportunity cost of cash and reduces investment in low-latency. However, the economic magnitude is small: a one standard deviation increase in endowment reduces speed investment by 3.33\% (0.02/0.060), and the effect is only statistically significant in some specifications. 

Table \ref{tab:tech} estimates the model \eqref{eq:main} for the three tiers of low-latency technology: that is, for high-cost, medium-cost, and low-cost as defined in Table \ref{tab:costs}. We find that traders are most sensitive to the magnitude of the speed bump if the marginal cost of low-latency technology is relatively low and they are better able to fine-tune their investment. For low cost technology, a one standard deviation increase in an asymmetric, deterministic message delay reduces investment by 16.67\% (0.10/0.60) compared to a drop of 8.95\% (1.8\%) for medium- (high-) cost speed technology. 

\begin{table}[H]
\caption{\label{tab:tech} \textbf{Low-latency investment and technology cost}}
    \centering
\vspace{0.1in}
\begin{tabular}{lccc} 
\toprule
& \multicolumn{3}{c}{Relative investment in speed} \\
 & (Low cost) & (Medium cost) & (High cost) \\
\cmidrule{2-4}
Constant & 0.60*** & 0.67*** & 0.54*** \\
 & (14.62) & (29.70) & (17.97) \\
Speed bump size ($\Delta$) & -0.10*** & -0.06* & -0.01 \\
 & (-3.02) & (-1.76) & (-0.23) \\
Symmetric speed bump ($d_\text{s}$) & 0.15** & 0.08 & 0.21*** \\
 & (2.38) & (1.65) & (3.25) \\
Random speed bump ($d_\text{r}$) & -0.02 & -0.01 & -0.02 \\
 & (-0.29) & (-0.31) & (-0.48) \\
Symmetric and random ($d_\text{s} \times d_\text{r}$) & -0.05 & -0.03 & 0.01 \\
 & (-0.71) & (-0.89) & (0.31) \\
Size $\times$ symmetric ($d_\text{s} \times \Delta$) & 0.07* & 0.04 & 0.00 \\
 & (1.95) & (1.61) & (0.03) \\
Size $\times$ random ($d_\text{r} \times \Delta$) & 0.05 & 0.02 & -0.02 \\
 & (1.37) & (0.52) & (-1.42) \\
Size $\times$ symmetric and random & -0.08* & 0.01 & -0.00 \\
 & (-2.06) & (0.23) & (-0.21) \\
Investment endowment & -0.03 & -0.02 & -0.00 \\
 & (-1.69) & (-0.95) & (-0.12) \\
Won in previous round & -0.02 & 0.01 & 0.03 \\
 & (-0.37) & (0.31) & (0.70) \\
\cmidrule{1-4}
Observations & 456 & 432 & 456 \\
R-squared & 0.28 & 0.21 & 0.21 \\
Participant FE & Yes & Yes & Yes \\
\bottomrule
\multicolumn{4}{c}{Robust t-statistics in parentheses. *** p$<$0.01, ** p$<$0.05, * p$<$0.1.} \\
\multicolumn{4}{c}{Two-way clustered standard errors (participant and period).} \\
\end{tabular}
\end{table}

Tables \ref{tab:risk_av} and \ref{tab:demographics_results} in Appendix \ref{sec:robust} estimate the model in \eqref{eq:main} for sub-samples of traders sorted by risk-aversion and demographic characteristics. We find that the main results are consistent across different sample splits.

\section{Conclusions} \label{sec:conclusion}
We study how investment in low-latency technology depends on the design of trading delays implemented by exchanges (or \emph{speed bumps}), in a laboratory experimental setting. We find that only asymmetric speed bumps deter investments in high speed technology. This is in line with the industry practice, since most exchanges implement asymmetric speed bumps, where liquidity taking orders are delayed and liquidity providing orders are not. 

Introducing an asymmetric speed bump reduces investment in low-latency technology by 20\%. The speed bump does not fully eliminate the arms' race, since speed investment is driven by competition between speculators as well as the competition between speculators and the market maker. A longer speed bump de-emphasizes competition between speculators and further reduces investment in speed technology. We find that implementing a symmetric speed bump leads to the same investment outcome as having no speed bump at all. Finally, we do not document significant differences between random and deterministic speed bumps.

Our results are of interest for exchange operators and financial regulators, in the context where multiple major trading venues either already implemented, or are planning to implement, order delays to curb the negative externalities of high-frequency trading data. Since technology investment data is highly proprietary, our experimental methodology sheds light on the optimal speed bump design and magnitude of its effect on the speed arms' race in financial markets.


\newpage

\bibliographystyle{jf}
\bibliography{references}

@article{Hagstromer2019InformationMarkets,
    title = {{Information Revelation in Decentralized Markets}},
    year = {Forthcoming},
    journal = {Journal of Finance},
    author = {Hagstr{\"{o}}mer, Bj\"{o}rn and Menkveld, Albert}
}

@ARTICLE{ChenFoley2017,
  author = {Chen, Haoming and Foley, Sean and Goldstein, Michael A. and Ruf, Thomas},
  title = {{The Value of a Millisecond: Harnessing Information in Fast, Fragmented Markets}},
  journal = {Working paper},
  year = {2017}
}

@ARTICLE{BrolleyCimon2018,
  author = {Brolley, Michael and Cimon, David},
  title = {{Order Flow Segmentation, Liquidity and Price Discovery: The Role of Latency Delays}},
  journal = {Working paper},
  year = {2018}
}

@article{FoucaultHombertRosu2015,
author = {Foucault, Thierry and Hombert, Johan and Ro\c{s}u, Ioanid},
title = {News Trading and Speed},
journal = {The Journal of Finance},
volume = {71},
year={2015},
number = {1},
pages = {335-382},
doi = {10.1111/jofi.12302},
url = {https://onlinelibrary.wiley.com/doi/abs/10.1111/jofi.12302},
eprint = {https://onlinelibrary.wiley.com/doi/pdf/10.1111/jofi.12302}
}

@article{MenkveldZoican2017,
author = {Menkveld, Albert J. and Zoican, Marius},
title = {Need for Speed? Exchange Latency and Liquidity},
journal = {The Review of Financial Studies},
volume = {30},
number = {4},
pages = {1188-1228},
year = {2017},
doi = {10.1093/rfs/hhx006},
URL = {http://dx.doi.org/10.1093/rfs/hhx006},
eprint = {/oup/backfile/content_public/journal/rfs/30/4/10.1093_rfs_hhx006/4/hhx006.pdf}
}

@article{bloomfield2015hidden,
  title={Hidden liquidity: Some new light on dark trading},
  author={Bloomfield, Robert and {O'Hara}, Maureen and Saar, Gideon},
  journal={The Journal of Finance},
  volume={70},
  number={5},
  pages={2227--2274},
  year={2015},
  publisher={Wiley Online Library}
}

@article{HoltLaury2002,
 ISSN = {00028282},
 URL = {http://www.jstor.org/stable/3083270},
 author = {Charles A. Holt and Susan K. Laury},
 journal = {The American Economic Review},
 number = {5},
 pages = {1644--1655},
 publisher = {American Economic Association},
 title = {Risk Aversion and Incentive Effects},
 volume = {92},
 year = {2002}
}

@Article{Aldrich2019,
author="Aldrich, Eric M.
and L{\'o}pez Vargas, Kristian",
title="Experiments in high-frequency trading: comparing two market institutions",
journal="Experimental Economics",
year="2019",
month="Mar",
day="08",
issn="1573-6938",
doi="10.1007/s10683-019-09605-2",
url="https://doi.org/10.1007/s10683-019-09605-2"
}

@Article{Crosetto2013,
author="Crosetto, Paolo
and Filippin, Antonio",
title="The ``bomb'' risk elicitation task",
journal="Journal of Risk and Uncertainty",
year="2013",
month="Aug",
day="01",
volume="47",
number="1",
pages="31--65",
issn="1573-0476",
doi="10.1007/s11166-013-9170-z",
url="https://doi.org/10.1007/s11166-013-9170-z"
}

@article{bloomfield2009noise,
  title={How noise trading affects markets: An experimental analysis},
  author={Bloomfield, Robert and {O'Hara}, Maureen and Saar, Gideon},
  journal={The Review of Financial Studies},
  volume={22},
  number={6},
  pages={2275--2302},
  year={2009},
  publisher={Society for Financial Studies}
}

@article{gozluklu2016pre,
  title={Pre-trade transparency and informed trading: Experimental evidence on undisclosed orders},
  author={Gozluklu, Arie E},
  journal={Journal of Financial Markets},
  volume={28},
  pages={91--115},
  year={2016},
  publisher={Elsevier}
}

@ARTICLE{Budish2015,
  author = {Budish, Eric and Cramton, Peter and Shim, John},
  title = {The High-Frequency Trading Arms Race: Frequent Batch Auctions as
	a Market Design Response},
  journal = {The Quarterly Journal of Economics},
  year = {2015},
  volume = {130},
  pages = {1547--1621}
}

@article{Aoyagi2019SpeedBumps,
    title = {{Speed Choice by High-Frequency Traders with Speed Bumps}},
    year = {2019},
    journal = {Manuscript},
    author = {Aoyagi, Jun}
}

@article{Aldrich2018OrderMessaging,
    title = {{Order protection through delayed messaging}},
    year = {2018},
    journal = {Manuscript},
    author = {Aldrich, Eric and Friedman, Daniel}
}

@article{Baldauf2019High-frequencyPerformance,
    title = {{High-frequency trading and market performance}},
    year = {2019},
    journal = {Manuscript},
    author = {Baldauf, Markus and Mollner, Joshua}
}

@Article{Shkilko2019,
  author =    {Shkilko, Andriy and Sokolov, Konstantin},
  title =     {{Every Cloud Has a Silver Lining: Fast Trading, Microwave Connectivity and Trading Costs}},
  journal =   {Journal of Finance},
  year =      {Forthcoming}
}

@article{baron2012trading,
  title={The Trading Profits of High Frequency Traders},
  author={Baron, Matthew and Brogaard, Jonathan and Kirilenko, Andrei},
  journal={Unpublished Manuscript},
  year={2012},
  publisher={Citeseer}
}

@article{brogaard2014high,
  title={High-Frequency Trading and Price Discovery},
  author={Brogaard, Jonathan and Hendershott, Terrence and Riordan, Ryan},
  journal={The Review of Financial Studies},
  volume={27},
  number={8},
  pages={2267--2306},
  year={2014},
  publisher={Oxford University Press}
}

@ARTICLE{Ye2013,
  author = {Ye, Mao and Yao, Chen and Gai, Jiading},
  title = {The Externalities of High Frequency Trading},
  journal = {Working paper},
  year = {2013}
}

@ARTICLE{HaasKhapkoZoican2018,
  author = {Haas, Marlene and Khapko, Mariana and Zoican, Marius},
  title = {Speed and Learning in High-Frequency Auctions},
  journal = {Working paper},
  year = {2018}
}

@article{PagnottaPhilippon2018,
author = {Pagnotta, Emiliano S. and Philippon, Thomas},
title = {Competing on Speed},
journal = {Econometrica},
volume = {86},
number = {3},
pages = {1067-1115},
year={2018},
doi = {10.3982/ECTA10762},
url = {https://onlinelibrary.wiley.com/doi/abs/10.3982/ECTA10762},
eprint = {https://onlinelibrary.wiley.com/doi/pdf/10.3982/ECTA10762}
}

@article{Hu2019IntentionalExchange,
    title = {{Intentional Access Delays, Market Quality, and Price Discovery: Evidence from IEX Becoming an Exchange}},
    year = {2019},
    journal = {Manuscript},
    author = {Hu, Edwin}
}

@article{biais2015equilibrium,
  title={Equilibrium Fast Trading},
  author={Biais, Bruno and Foucault, Thierry and Moinas, Sophie},
  journal={Journal of Financial economics},
  volume={116},
  number={2},
  pages={292--313},
  year={2015},
  publisher={Elsevier}
}

@ARTICLE{AndersonWalton2018,
  author={Lisa Anderson and Emad Andrews and Baiju Devani and Michael Mueller and Adrian Walton},
  title={{Speed Segmentation on Exchanges: Competition for Slow Flow}},
  year={2018},
  journal={Bank of Canada Staff Working Papers},
  volume={18-3},
  doi={},
}

@article{KyleLee2017,
author = {Kyle, Albert S and Lee, Jeongmin},
title = {Toward a fully continuous exchange},
journal = {Oxford Review of Economic Policy},
volume = {33},
number = {4},
pages = {650-675},
year = {2017},
doi = {10.1093/oxrep/grx042},
URL = {http://dx.doi.org/10.1093/oxrep/grx042},
eprint = {/oup/backfile/content_public/journal/oxrep/33/4/10.1093_oxrep_grx042/2/grx042.pdf}
}

@article{ChenSchonger2016,
	Author = {Chen, Daniel L. and Schonger, Martin and Wickens, Chris},
	Da = {2016/03/01/},
	Doi = {https://doi.org/10.1016/j.jbef.2015.12.001},
	Isbn = {2214-6350},
	Journal = {Journal of Behavioral and Experimental Finance},
	Pages = {88--97},
	Title = {oTree---An open-source platform for laboratory, online, and field experiments},
	Ty = {JOUR},
	Url = {http://www.sciencedirect.com/science/article/pii/S2214635016000101},
	Volume = {9},
	Year = {2016}}

 \newpage
 \appendix
 \singlespacing

 \numberwithin{equation}{section}
 \numberwithin{prop}{section}
 \numberwithin{lem}{section}
 \numberwithin{defn}{section}
 \numberwithin{cor}{section}
 \numberwithin{figure}{section}
 \numberwithin{table}{section}

\section{Robustness checks \label{sec:robust}}

\begin{table}[H]
\caption{\label{tab:risk_av} \textbf{Low-latency investment by participant risk-aversion}}
    \centering
\vspace{0.1in}
\begin{tabular}{lcccc} 
\toprule
& \multicolumn{4}{c}{Relative investment in speed} \\
 & \multicolumn{2}{c}{Bomb task risk aversion} & \multicolumn{2}{c}{\citet{HoltLaury2002} risk aversion} \\
 & (High) & (Low) & (High) & (Low) \\
\cmidrule{2-5}
Constant & 0.58*** & 0.62*** & 0.64*** & 0.63*** \\
 & (22.82) & (28.89) & (20.70) & (17.07) \\
Speed bump size ($\Delta$) & -0.06 & -0.05* & -0.05* & -0.04 \\
 & (-1.58) & (-1.97) & (-1.90) & (-0.91) \\
Symmetric speed bump ($d_\text{s}$)  & 0.21*** & 0.09** & 0.10** & 0.15 \\
 & (3.66) & (2.23) & (2.46) & (1.73) \\
Random speed bump ($d_\text{r}$) & 0.00 & -0.03 & -0.03 & -0.00 \\
 & (0.10) & (-0.84) & (-0.82) & (-0.11) \\
Symmetric and random ($d_\text{s} \times d_\text{r}$) & -0.07 & 0.02 & -0.02 & -0.03 \\
 & (-1.75) & (0.46) & (-1.07) & (-0.33) \\
Size $\times$ symmetric ($d_\text{s} \times \Delta$) & 0.07* & 0.00 & 0.04 & 0.02 \\
 & (1.82) & (0.06) & (1.51) & (0.26) \\
Size $\times$ random ($d_\text{r} \times \Delta$) & 0.04 & -0.00 & 0.04* & -0.01 \\
 & (0.79) & (-0.06) & (1.82) & (-0.14) \\
Size $\times$ symmetric and random & -0.07 & 0.01 & -0.05** & 0.03 \\
 & (-1.23) & (0.65) & (-2.16) & (0.90) \\
Investment endowment & -0.02 & -0.01 & -0.01 & -0.04 \\
 & (-1.72) & (-0.80) & (-0.75) & (-1.59) \\
Won in previous round & 0.05** & -0.04 & -0.03 & 0.04 \\
 & (2.80) & (-1.39) & (-0.98) & (0.92) \\
\cmidrule{1-5}
Observations & 648 & 696 & 720 & 384 \\
R-squared & 0.20 & 0.26 & 0.26 & 0.21 \\
Participant FE & Yes & Yes & Yes & Yes \\
 Group FE & Yes & Yes & Yes & Yes \\ 
\bottomrule
\multicolumn{5}{c}{Robust t-statistics in parentheses. *** p$<$0.01, ** p$<$0.05, * p$<$0.1.} \\
\multicolumn{5}{c}{Three-way clustered standard errors (participant, group, and period).} \\
\end{tabular}
\end{table}

\begin{table}[H]
\caption{\label{tab:demographics_results} \textbf{Impact of speed bumps by participant demographics}}
    \centering
\vspace{0.1in}

\begin{tabular}{lllllll} 
\toprule
& \multicolumn{6}{c}{Relative investment in speed} \\
& \multicolumn{2}{c}{Finance courses} & \multicolumn{2}{c}{Trading experience} & \multicolumn{2}{c}{Gender} \\
 & No & Yes & Yes & No & Male & Female \\
\cmidrule{2-7}
Constant & 0.58*** & 0.62*** & 0.60*** & 0.60*** & 0.57*** & 0.62*** \\
 & (15.81) & (25.36) & (12.26) & (45.30) & (15.82) & (30.65) \\
Speed bump size ($\Delta$) & -0.08* & -0.04* & -0.03 & -0.06** & -0.09** & -0.04 \\
 & (-2.03) & (-2.07) & (-0.95) & (-2.38) & (-2.34) & (-1.57) \\
Symmetric speed bump ($d_\text{s}$) & 0.18** & 0.12** & 0.11 & 0.16*** & 0.18** & 0.13*** \\
 & (2.91) & (2.69) & (1.45) & (4.93) & (2.27) & (4.76) \\
Random speed bump ($d_\text{r}$) & -0.02 & -0.02 & -0.08 & 0.01 & 0.00 & -0.02 \\
 & (-0.51) & (-0.52) & (-1.41) & (0.36) & (0.05) & (-0.71) \\
Symmetric and random ($d_\text{s} \times d_\text{r}$) & -0.02 & -0.02 & 0.11* & -0.07*** & 0.04 & -0.06** \\
 & (-1.70) & (-0.50) & (2.16) & (-4.24) & (0.97) & (-2.45) \\
Size $\times$ symmetric ($d_\text{s} \times \Delta$)  & 0.08** & 0.00 & -0.01 & 0.05** & 0.02 & 0.05** \\
 & (2.72) & (0.16) & (-0.15) & (2.31) & (0.52) & (2.17) \\
Size $\times$ random ($d_\text{r} \times \Delta$) & 0.01 & 0.02 & 0.05** & 0.01 & 0.02** & 0.02 \\
 & (0.26) & (1.06) & (2.23) & (0.21) & (2.13) & (0.41) \\
Size $\times$ symmetric and random & -0.02 & -0.03*** & -0.07*** & -0.01 & -0.02*** & -0.03 \\
 & (-0.44) & (-6.18) & (-3.41) & (-0.35) & (-4.24) & (-0.77) \\
Investment endowment & -0.02 & -0.02 & -0.03 & -0.02 & -0.01 & -0.02 \\
 & (-1.38) & (-1.14) & (-1.02) & (-1.40) & (-0.62) & (-1.59) \\
Won in previous round & 0.05 & -0.03 & 0.01 & -0.00 & 0.05 & -0.02 \\
 & (1.14) & (-1.00) & (0.55) & (-0.20) & (1.42) & (-0.89) \\
\cmidrule{1-7}
Observations & 552 & 792 & 336 & 1,008 & 480 & 864 \\
R-squared & 0.23 & 0.23 & 0.23 & 0.23 & 0.23 & 0.25 \\
Participant FE & Yes & Yes & Yes & Yes & Yes & Yes \\
 Group FE & Yes & Yes & Yes & Yes & Yes & Yes \\ 
 \bottomrule
\multicolumn{7}{c}{Robust t-statistics in parentheses. *** p$<$0.01, ** p$<$0.05, * p$<$0.1.} \\
\multicolumn{7}{c}{Three-way clustered standard errors (participant, group, and period).} \\
\end{tabular}
\end{table}

\newpage

\section*{Experimental instructions \label{sec:instructions}}

\subsection*{Welcome!}

Welcome and thank you for participating in the experiment. 

\noindent Please keep silent and remain seated at your place. Please note that during the experiment any communication with other participants is not allowed. Please switch off your mobile phones. Should you have any questions, raise your hand and an experimenter will answer you privately. All the decisions are made anonymously and no participant under any circumstances will know the identity of any other participant.

\noindent In addition to receiving course credit for completing the experiment, you will be paid your earnings from the experiment by means of an Amazon.ca eGift Card. After you are finished, the computer will select one of the trading rounds at random. This will be your payment round: You will receive the profit that you made in that round. You should treat every round of the experiment as if it is the one that counts, because it might be!

\subsection*{Who is trading?}

\begin{itemize}
\item You will be assigned to groups of 3 traders. 

\item Each of the 3 traders in your group will be trading against a computer. You will not be trading with each other.

\item The experiment consists of 32 rounds of trading. Each of the rounds will have different trading conditions which will be explained to you on your computer screen.  

\item At the beginning of each round, you receive an endowment of ECoins to trade. ECoins are a virtual currency. The exchange rate between ECoins (laboratory dollars) and Canadian dollars is: 1 ECoin = 0.5 CAD\$. You start with a new endowment each round.

\end{itemize}

\subsection*{What are you trading?}

\begin{itemize}
\item Each round, you will face a profitable trading opportunity worth 100 ECoins  on a laboratory market.

\item What exactly does this mean? You can think about it as follows. At the beginning of each round, the computer makes a mistake. It sells an asset worth 100 ECoins for a price of zero. Therefore, there is a profitable trading opportunity worth 100 ECoins. To take advantage of the opportunity, you submit a trade buy order to the market.




\end{itemize}

\subsection*{How does trading work?}

\begin{itemize}

\item In order to win in each trading round and earn the profit, your order needs to reach the market 
before the orders of your two competitors.
\item How can you become faster than other traders in your group? 
\item Each round you decide how much of your endowment of ECoins to invest in speed. Your investment in speed will determine how quickly you can arrive at the market to trade. You do not have to spend the full amount: Anything that you do not spend will be returned to you as part of the round payoff.
\item Orders take a random time to arrive to the market (think network lag), and are executed sequentially: first come, first served. 
\item Further, the computer can cancel her quote (in which case the profit opportunity disappears).
\item Therefore, to trade and earn the profit your order needs to arrive before any other traders' in the group \textbf{and} before the computer cancels the quote.
\end{itemize}

\subsection*{Beware of ``speed bumps"!}

\begin{itemize}
\item In some rounds, the market is slowed down by a trading delay, also called a \textbf{speed bump}.
\item With a speed bump, yours and other traders' orders are delayed by a certain amount of time and will take longer to be executed.
\item In some rounds, the computer is also affected by the speed bump: her decision to cancel orders is delayed by the same amount.
\item Hence, the speed bump can be \emph{symmetric} or \emph{asymmetric}. In case of a symmetric speed bump everyone gets delayed: all traders in your group and the computer. In case of an asymmetric speed bump only the human traders (traders in your group) are delayed.
\item The speed bump you face can be \emph{deterministic} (or fixed). For example, a five second delay. Alternatively, the speed bump can be \emph{random}. What does it mean? For example, you will know that you  are equally likely to be delayed by either 0.5 seconds, or 1.0 seconds, or 1.5 seconds. 
\item You will be informed in each round about the type and the length of the speed bump (if any).
\end{itemize}

\subsection*{Example of the speed investment decision}
\begin{itemize}
\item Each participant will be informed in each trading round on
\begin{itemize}
    \item[(a)] The endowment in ECoins that can be invested in speed technology.
    \item[(b)] The type of the speed bump traders are facing in this round: How long is the delay? Fixed or random? Who is affected? 
\end{itemize}

\item You will also see calculations and a visual demonstration that show you how different speed investments affect how fast you can expect your order to be executed.

\item Based on this information you need to make a decision: How much do you want to invest in speed in this round? 

\item In the example below, your choice is to invest in speed 15.25 out of available 20 ECoins.
\end{itemize}
\begin{figure}[H]
\begin{centering}
\includegraphics[width=\textwidth]{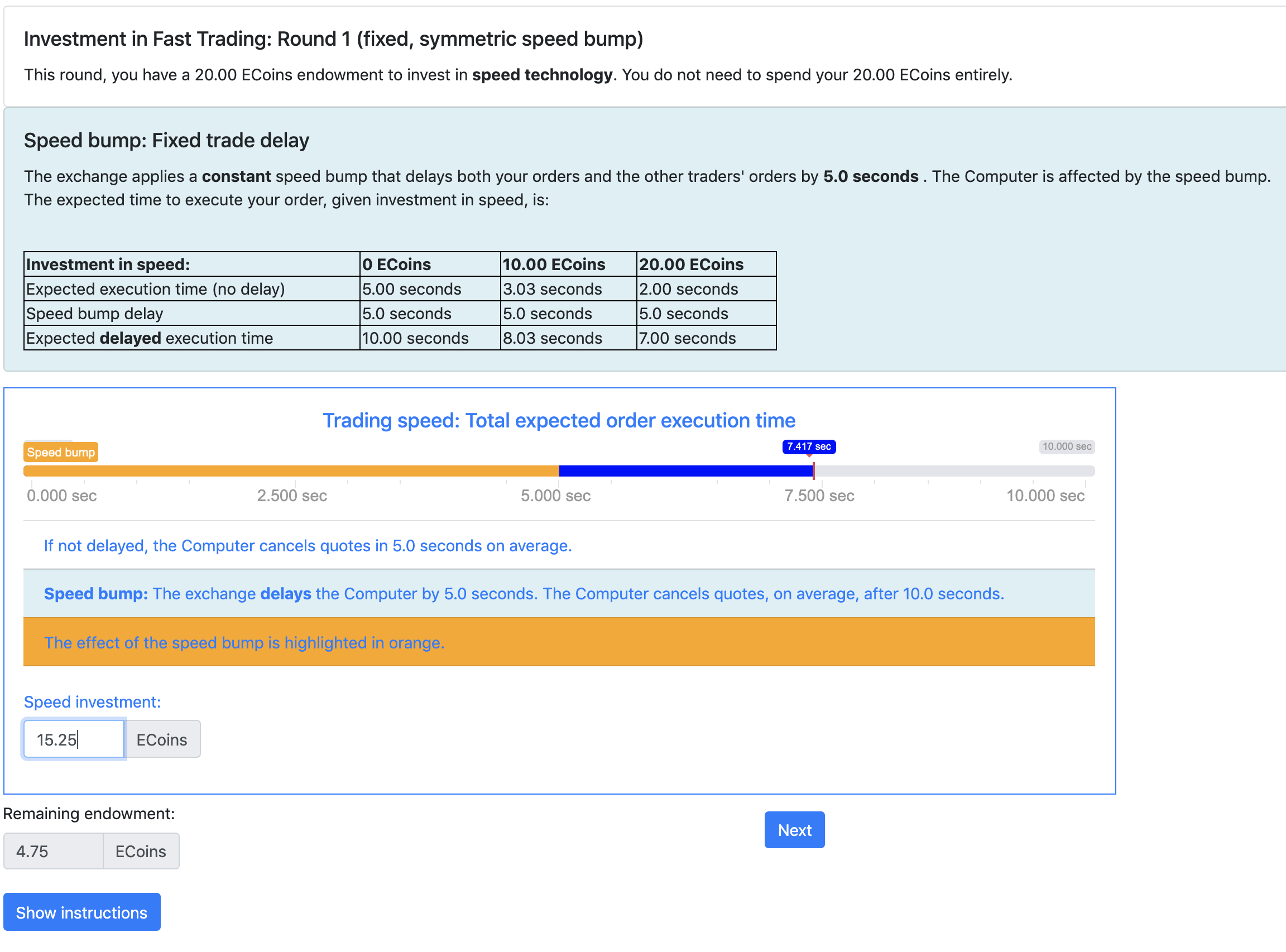}
\par\end{centering}
\end{figure}

\subsection*{Example of order execution}

\begin{itemize}
\item After all traders in your group made their investment in speed, everyone's orders are sent to the market at the same time. 
\item You will see a trading clock (up to millisecond precision), an indicator on whether the trading opportunity is still available or not, and a market activity panel with order execution.
\item In the example below, your order reached the market first (highlighted in green).
\end{itemize}

\begin{figure}[H]
\vspace{0.1in}
\begin{centering}
\includegraphics[width=\textwidth]{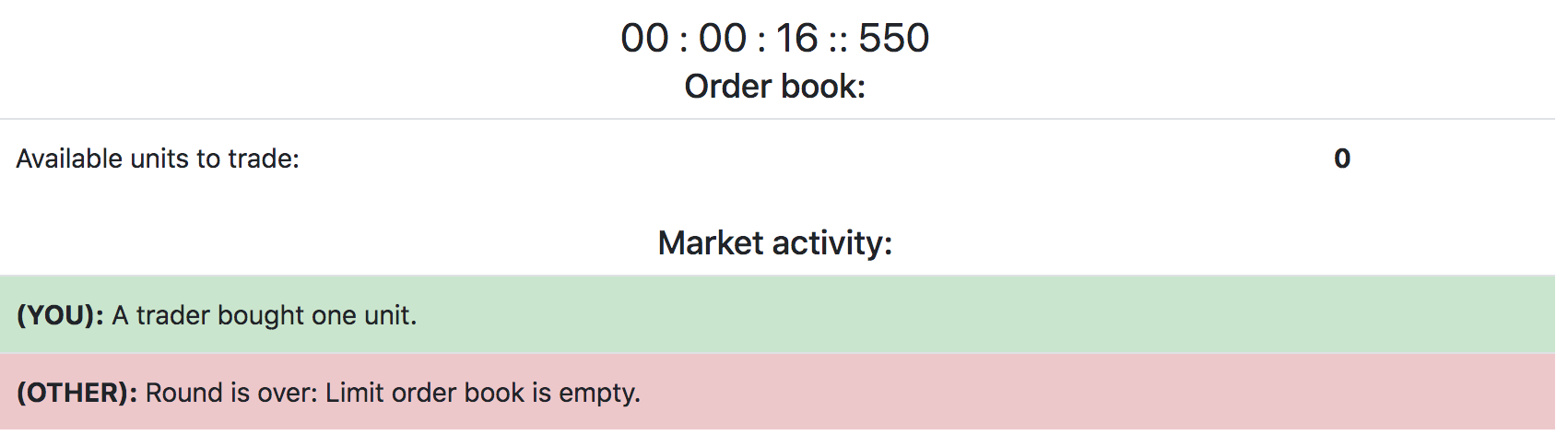}
\par\end{centering}
\end{figure}

\subsection*{Example of end-of-round payoffs}
\begin{itemize}
    \item After each round, you will see a ``Results" screen confirming whether your order was executed and your payoff for the round.
    \item In the example below, you invested 3 out of 20 available ECoins in speed, you were the first to reach the market, and your order was executed. Therefore, your profit for this round is 100+17=117 ECoins. 
\end{itemize}

\begin{figure}[H]
\begin{centering}

\includegraphics[width=\textwidth]{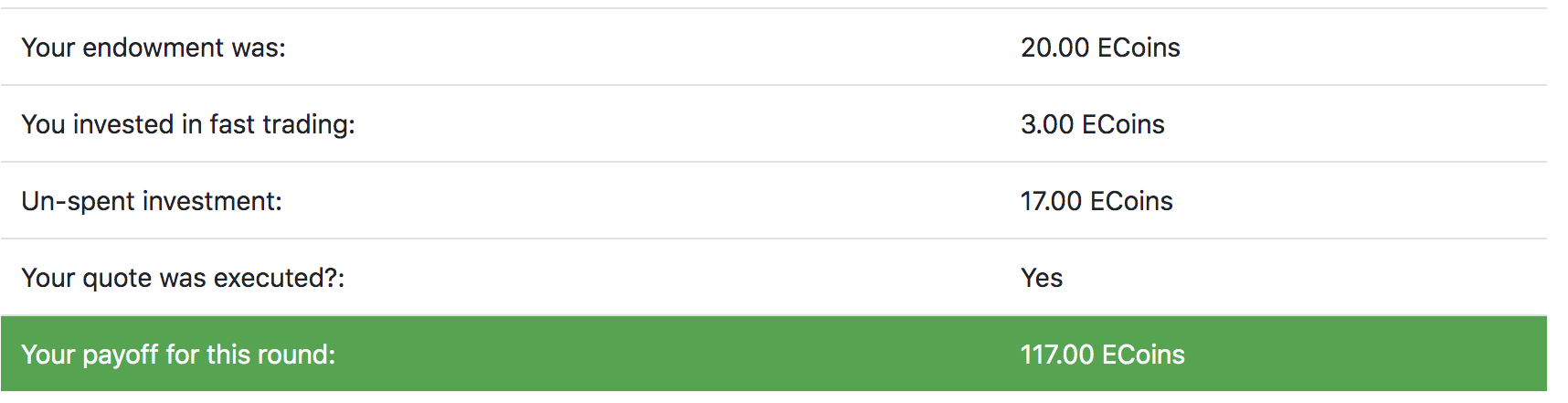}

\par\end{centering}
\end{figure}

\subsection*{At the end}

After the 32 trading rounds you will be asked to answer some questions on the computer. All information will remain confidential.

\bigskip

\begin{shadedbox}
\textbf{Summary}
\begin{enumerate}
\item In each trading round, you will trade against a computer competing with two other human traders for a profit opportunity worth 100 ECoins.
\item To earn the 100 ECoins profit you need your trade order to arrive at the market fast: before other traders in your group and before the computer cancels the profit opportunity.
\item You will have a sum of ECoins in each trading round that you can invest to become faster. High investment in trading speed increases your chances to arrive at the market first and trade, but does not guarantee it.
\item After all traders in your group choose their speed investments, everyone's orders are sent to the market. You will see the market activity information on your screen and you will know if you managed to capture the trading opportunity or not.
\item There will be 32 trading rounds. In each round you can decide on how much to invest in trading speed and have the opportunity to earn profit. Each trading round will differ in the trading conditions you and your competitors face.
\item The most important difference between the trading rounds you need to pay attention to is the trading delay, or the speed bump, imposed by the exchange.
\item At the end of your experiment, one of the trading rounds will be selected at random as your payment round.  So think and play in each round as if it is the round that counts, because it might be!
\end{enumerate}
\end{shadedbox}

\end{document}